\documentclass[%
 ,onecolumn%
 ,secnumarabic%
,showpacs,amssymb, amsmath,nobibnotes, aps, prd]{revtex4}

\usepackage{bm}%
\usepackage{epsfig}
\expandafter\ifx\csname package@font\endcsname\relax\else
 \expandafter\expandafter
 \expandafter\usepackage
 \expandafter\expandafter
 \expandafter{\csname package@font\endcsname}%
\fi

\begin{document}
\title{Gravitational wave burst search in the Virgo C7 data}%

\author{F. Acernese$^{5ac}$,
M. Alshourbagy$^{10ab}$,
F. Antonucci$^{11a}$,
S. Aoudia$^{6}$,
K. G. Arun$^{7}$,
P. Astone$^{11a}$,
G. Ballardin$^2$,
F. Barone$^{5ac}$,
L. Barsotti$^{10ab}$,
M. Barsuglia$^{14}$
% ...
}

\address{$^1$Laboratoire d'Annecy-le-Vieux de Physique des Particules (LAPP),  IN2P3/CNRS, Universit\'e de Savoie, F-74941 Annecy-le-Vieux, France}
\address{$^2$European Gravitational Observatory (EGO), I-56021 Cascina (Pi), Italia}
\address{$^3$INFN, Sezione di Firenze, I-50019 Sesto Fiorentino$^a$; Universit\`a degli Studi di Firenze, I-50121$^b$, Firenze;  Universit\`a degli Studi di Urbino "Carlo Bo", I-61029 Urbino$^c$, Italia}
\address{$^4$Laboratoire des Mat\'eriaux Avanc\'es LMA, IN2P3/CNRS, 
F-69622 Villeurbanne, Lyon, France}
\address{$^5$ INFN, sezione di Napoli $^a$; Universit\`a di Napoli "Federico II"$^b$ Complesso Universitario di Monte S.Angelo, I-80126 Napoli; Universit\`a di Salerno, Fisciano, I-84084 Salerno$^c$, Italia}
\address{$^{6}$ Departement Artemis,  Observatoire de la C\^ote d'Azur, CNRS, F-06304 Nice,  France.}
\address{$^{7}$LAL, Universit\'e Paris-Sud, IN2P3/CNRS, F-91898 Orsay, France}
\address{$^{8}$ESPCI, CNRS,  F-75005 Paris, France}
\address{$^{9}$INFN, Sezione di Perugia$^a$; Universit\`a di Perugia$^b$, I-6123 Perugia,Italia}
\address{$^{10}$INFN, Sezione di Pisa$^a$; Universit\`a di Pisa$^b$; I-56127 Pisa; Universit\`a di Siena, I-53100 Siena$^c$, Italia}
\address{$^{11}$INFN, Sezione di Roma$^a$; Universit\`a "La Sapienza"$^b$, I-00185  Roma, Italia}
\address{$^{12}$National institute for subatomic physics, NL-1009 DB$^a$;  Vrije Universiteit, NL-1081 HV  $^b$, Amsterdam, The Netherlands}
\address{$^{13}$INFN, Sezione di Roma Tor Vergata$^a$; Universit\`a di Roma Tor Vergata$^b$, Istituto di Fisica dello Spazio Interplanetario (IFSI) INAF$^c$, I-00133 Roma; Universit\`a dell'Aquila, I-67100 L'Aquila$^d$, Italia}
\address{$^{14}$ AstroParticule et Cosmologie (APC), CNRS: UMR7164-IN2P3-Observatoire de Paris-Universit\'e Denis Diderot-Paris VII - CEA : DSM/IRFU}
\address{$^*$Permanent address: School of Physics \& Astronomy University of Birmingham B15 2TT, UK}


\date{\today ~~ - DRAFT 5.4.2}%

\begin{abstract}
A search for gravitational wave burst events has been performed with the Virgo C7 commissioning run data that have
been acquired in September 2005 over five days.
It focused on un-modeled short duration signals in the frequency range 150 Hz to 2 kHz.
A search aimed at detecting the GW emission from the merger and ringdown phases of binary black hole coalescences was also
carried out.
An extensive understanding of the data was required to be able to handle a burst search using the output of only one detector.
A 90\% confidence level upper limit on the number of expected events given the Virgo C7 sensitivity
curve has been derived as a function of the signal strength, for un-modeled gravitational wave search.
The sensitivity of the analysis presented is, in terms of the root sum square strain amplitude, $h_{rss} \simeq  10^{-20} / \sqrt{Hz}$.
This can be interpreted in terms of a frequentist upper limit on the rate ${\cal{R}}_{90\%}$ of detectable gravitational 
wave bursts at the level of 1.1 events per day at 90\% confidence level. 
From the binary black hole search, we obtained the distance reach at 
50\% and 90\% efficiency as a function of the total mass of the final black hole. 
The maximal detection distance for non-spinning high and equal mass black hole 
binary system obtained by this analysis in C7 data is $\simeq$ 2.9 $\pm$ 0.1 Mpc for 
a detection efficiency of 50\% for a binary of total mass $80 M_{\odot}$.
\end{abstract}

\maketitle

\section{Introduction}

Virgo \cite{ref:virgo_recent} is a 3-km long arm power-recycled Michelson interferometer located near Pisa, Italy, 
whose goal is to detect gravitational waves (GW) emitted by astrophysical sources extending out past the Virgo cluster. 
The commissioning of the detector started in 2003 and regular data taking campaigns have been organized after each 
important milestone.
The last commissioning run (C7) took place in September 2005 and lasted for 5 days.
The best achieved sensitivity was $h \simeq 7 \times 10^{-22}/\sqrt{Hz}$ at 300 Hz.
The Virgo design sensitivity at this frequency is expected to be better by an order of magnitude assuming that 10 W enters
into the interferometer. 
However, during the C7 run, Virgo was running with a reduced light power, 0.7 W, because the backscattering in the 
mode-cleaner cavity of the light reflected by the recycling mirror prevented the control of the interferometer at full power.
Despite the reduced sensitivity, several GW searches have been carried out using this data set in order to set up and tune 
the pipelines developed in Virgo: the search of the coalescence of neutron stars \cite{ref:C7_CB}, 
the search of continuous wave sources \cite{ref:C7_pulsar}, a joint Virgo, AURIGA, EXPLORER and NAUTILUS GW search 
\cite{ref:C7_bars}, a targeted GW burst search associated with the long GRB 050915a \cite{ref:C7_GRB}, and the search of 
GW burst signals that is reported in this paper.

The category of gravitational wave bursts includes all possible signals whose duration is short, less than a few hundreds of 
milliseconds.
Many violent astrophysical phenomena will be accompanied by an emission of GW burst.
If these events happen sufficiently close by, the ground based gravitational detectors will be able to observe these 
short duration GW bursts.
There are many candidate GW burst sources.
These include massive star core collapse \cite{ref:zwerger, ref:dimmelmeier02, ref:ott04, ref:shibata, ref:ott06}, the merging 
phase of coalescing compact binary systems forming a single black hole (BH) \cite{ref:flanagan, ref:baker, ref:pretorius, 
ref:campanelli}, 
BH ring-down \cite{ref:kokkotas}, astrophysical engines that generate gamma-ray bursts (GRB) \cite{ref:meszaros}, 
neutron star oscillation modes and instabilities \cite{ref:ferrari}, or cosmic string cusps and kinks \cite{ref:damour}.
Some of these sources are well modeled, but not all, and as such a burst pipeline is built by 
making very few assumptions about the nature of these waveforms.
The theoretical event rates for many of these sources are quite uncertain.

Binary black hole (BBH) mergers are an interesting source for GW burst searches. 
The typical event rate for these sources is 1 $\mathrm{Myr}^{-1}$ per galaxy \cite{ref:postnov06,ref:oshaughnessy05} 
and is still highly uncertain given the lack of direct observational evidence for the existence of BBH systems, 
unlike the double neutron star systems.
Though these sources are routinely searched for by the inspiral phase prior to merger using matched filtering techniques,
an independent search focusing on the merger and ring-down GW emission, which is dominant for high mass binaries 
\cite{ref:flanagan}, may bring additional confidence \footnote{Template based searches for binary black hole sources 
using different waveforms which capture the different phases, including merger and ringdown, 
are now applied to analyze the LIGO and Virgo data.}.
A search of this type was first attempted with the LIGO S2 data~\cite{ref:LIGO_burst_S2};
this used the then-available numerical relativity waveforms from the Lazarus project
as a model of BBH mergers. However the estimates obtained were projected as order of 
magnitude estimates due to the nonrobusteness of the waveforms used.
Though the burst searches do not have as large of a distance reach as an inspiral search, it will be interesting
to understand the implications of a BH binary search from a burst perspective, especially because a burst search is sensitive 
to the merger and ring-down phase of a BBH coalescence.
Among the three phases of inspiral, merger and ring-down, most of the energy may be released in the highly relativistic phase of 
merger, that is difficult to model by analytic approximation methods. 
But recently numerical relativity simulations have made significant advances in generating the waveforms for all the three phases 
of the binary evolution (see \cite{ref:pretorius07} and references there in).

In this paper, we report on an all-sky burst search for un-modeled waveforms and BBH mergers without using any prior 
information on the expected waveform.
This concerns all short duration ($\ll$ 1 s) signals having energy in the best sensitivity frequency band of the C7 data,
150-2000 Hz.
This search has been performed using the data of only one interferometer which does not allow one to perform timing coincidence 
to eliminate spurious events in one of the detectors, as was done in the burst searches performed using the LIGO detectors 
\cite{ref:LIGO_burst_S1, ref:LIGO_burst_S2, ref:LIGO_burst_S4} and
the bar detectors \cite{ref:IGEC_burst}.
However, a similar one-detector burst search has already been carried out with the data from the TAMA detector \cite{ref:TAMA_burst}.
The C7 run data has been used for extensive analysis of the Virgo noise; this is a fundamental step in the path toward performing a 
GW burst search with only one detector. 
Indeed, numerous sources of noise generate, in the interferometer's GW strain amplitude channel, transient events which mimic GW burst events; 
environmental noise, such as acoustic noise, has been found at the origin of many of them. 
Non stationary data can also generate an excess of short duration events. 
Such phenomena have been found, examined, and understood in the C7 data.
These studies allowed us to define the pre-processing and post-processing steps needed in order
to optimize the performance of the GW burst 
pipeline used in this analysis.

The outline of this paper is as follows.
In Section \ref{sec:section2} we describe the main features of the Virgo detector, emphasizing the operating characteristics
during the C7 run
that played a role in understanding the quality of the data.
In Section \ref{sec:dataset} we describe how we select the data periods that have been used for 
this analysis.
The burst pipeline used in this analysis is described in Section \ref{sec:section3}.
In Section \ref{sec:DQ} the main results on the C7 data characterization needed in order to
understand and to suppress the sources of glitches are summarized;
this includes the veto strategy used against identified sources of noise.
Section \ref{sec:background} gives the results of the search in the C7 data.
Section \ref{sec:section5} explains how the analysis sensitivity has been estimated considering different types of possible 
GW burst waveforms. 
We especially considered in this paper BBH merger and ring-down numerical waveforms \cite{ref:goddard}. 
Section \ref{sec:section6} gives the upper limits obtained at 90\% of confidence level on the number of events as a function of the signal strain amplitude. 
We finally conclude this paper with an astrophysical interpretation of the present GW burst search.

\section{Virgo during the Commissioning run C7}
\label{sec:section2}

\subsection{Detector status}
\label{sec:detector}
The Virgo detector \cite{ref:virgo} is a power recycled Michelson interferometer with 3-km long arms that each 
contain a Fabry-Perot 
cavity.
All mirrors are suspended from the so-called Superattenuator \cite{ref:SA}, whose goal is to reduce drastically 
above 10 Hz the seismic noise 
transferred to the instrument.
A 20 W Nd:YAG laser is used to illuminate the interferometer.
The laser light is modulated in phase at the frequencies $\simeq $ 22 and 6\,MHz; this technique permits the GW strain to be 
detected at the modulation frequency where the laser power fluctuation is much smaller than in the interferometer
bandwidth.
The beam is spatially filtered with a 144\,m long triangular input mode-cleaner cavity before being injected 
into the main interferometer.
The laser frequency is pre-stabilized in order to acquire the control of the different optical cavities, but to reach the 
extreme sensitivity targeted by Virgo an enhanced control of the laser frequency noise is required; 
it has to be reduced by several orders of magnitude.
This is the role of the so-called {\it second stage frequency stabilization} which is engaged during 
the cavities' lock acquisition \cite{ref:locking}.
The beam entering the interferometer is divided by the beam splitter (BS) into two beams that are injected into the 3-km long 
arm cavities.
Apart from the mirrors' losses, all light fed-in by the injection system subsequently returns to it. 
The power-recycling (PR) mirror, with a reflectivity of 92\%, reflects the out-going light back to the main interferometer.
Together with the Michelson interferometer the power-recycling mirror forms a Fabry-Perot-like cavity 
in which the light power is resonantly enhanced, thereby improving the shot noise limit. 
The Michelson interferometer is held on the dark fringe, and the GW strain signal is expected in the 
beam at the dark port, which leaves the vacuum via the so-called detection bench.
The detection bench is a suspended table accommodating several optical components. 
The beam coming from the BS passes through an output mode-cleaner, a 3.6 cm long rigid cavity.
The main output beam is detected by a pair of InGaAs photodiodes.
Useful signals are obtained by detecting the light in the transmission of the arm cavities and in reflection of the 
power-recycling cavity (B2 photodiode). 
The GW signal that results from a detuning of the carrier resonance in the arms is extracted from 
the dark port channel, demodulated, and then sampled at 20 kHz.
This signal is digitized and filtered. 

The control of the interferometer consists in maintaining the laser light resonant in the optical cavities and the output 
port tuned on the dark fringe, defining its working point.
More precisely, the carrier must be resonant in all cavities while the sidebands must be resonant in the central cavity
but anti-resonant in the arms.
Despite the good seismic noise attenuation provided by the Superattenuator, feedback controls are mandatory in order to keep
the interferometer locked on the right working point. 
In addition to the control of the longitudinal degrees of freedom of the cavities, the mirrors must be kept aligned with 
respect to each other. This is the so-called auto-alignment control (AA).
If uncontrolled, the angular degrees of freedom of the suspended optics distort the cavity eigen-modes.
This causes power modulation of the light fields;
furthermore, long term drifts will make the longitudinal control impossible after a certain amount of time, 
and misalignments increase the coupling of other noise sources into the dark port. This has been a major problem in the
C7 data analysis (see Section \ref{sec:DQ}). 
In actuality, during C7 not all of the mirrors were under AA control; the arm input mirrors and the injection bench were only 
controlled locally.

To convert the signal received on the dark port into a measure of GW strain one needs to calibrate the residual 
arm-length difference; this includes at low frequencies the effect of the control loop keeping the cavities in resonance.
The loop correction signals are subtracted from the dark port signal that is then converted into a strain via knowledge of 
the optical gain of the feedback control loops.
This gain may vary depending, for instance, on the alignment drifts of the optical elements.
To monitor the variation, some sinusoidal signals (four between 100 and 110 Hz, and four between 350 and 360 Hz) 
are applied on the end-arm mirrors and the amplitude of the lines are measured.
Another important aspect of the GW strain reconstruction is the good knowledge of the transfer function of the actuator 
chain which is used to control the mirrors.
Actually, during the C7 run, the calibration procedure was not as accurate as it is now. For instance, 
the frequency dependence of the digital to analog converter transfer function was not taken into account 
in the reconstruction of the GW strain amplitude channel $h(t)$.
This yielded a systematic error on $h(t)$ which has been estimated to be 40\% \cite{ref:hrec_C7}.

\begin{figure}
\begin{center}
\epsfig{file=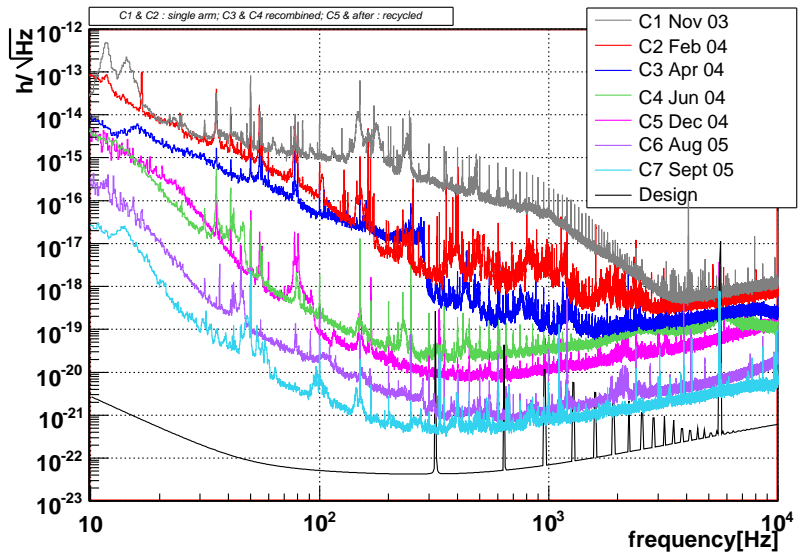, height=8cm}
\caption{\label{fig:sensitivity} Sensitivity curves obtained between 2003 and 2005 during the commissioning of the Virgo detector 
before the 2005 shutdown.
The black curve is the Virgo nominal sensitivity curve which has been computed assuming a 10 W laser. }
\end{center}
\end{figure}

Moreover, during C7 Virgo was running with a 700 mW laser beam entering the interferometer in order to avoid 
instabilities due to the backscattering of light into the mode-cleaner cavity reflected by the recycling mirror.
The problem has since been fixed during a shutdown carried out just after the C7 run. 
This reduced power light limited the sensitivity at high frequencies.
FIG. \ref{fig:sensitivity} displays and compares the progress achieved on the sensitivity curve since the 
beginning of the Virgo commissioning.
At high frequencies ($>$ 300 Hz), the sensitivity was limited by the shot noise and the laser frequency noise. 
At low frequencies ($<$ 100 Hz) the longitudinal and angular controls of the mirrors are the main limitation;
the electronic noise of the actuator and/or the sensor introduced by the control loops induce a displacement 
of the mirror which is well above the fundamental noise floor.

\subsection{C7 data set}
\label{sec:dataset}
After the lock acquisition is completed, the interferometer state is set to {\it science mode}.
The science mode duty cycle maintained by Virgo during C7 was 66\%. The data was split into 53 data blocks, 
or segments, spanning over the 5 days. 
A minimal duration of 10 minutes has been required for a segment.
In order to define the start and end time of the data segments of good quality for the search for burst GW events we considered 
all the known instrumental effects that influence and spoil the output of the burst pipeline. For example, the last 10 seconds of each segment 
has been removed because many signals, including the GW strain channel, start oscillating and thereby cause the loss of lock.
Another problem concerns the excitation of some thermal resonances of the last stage of the suspension (violin modes) 
when attempting to lock the interferometer. 
The decay time of these resonances has been determined, and for each segment we suppressed the beginning until the 
amplitude of the resonance having the longest decay time has been decreased by 90\%.
The reduction of the size of the data segments (hereafter called {\it science segments}) due to these data quality criteria 
amounts to 8\%. 
In addition, some signals have been artificially injected into the GW strain channel during C7 by applying
a force on the input mirror of one of the interferometer arm's Fabry-Perot cavity. These are the so-called hardware injections.
The hardware injection signals are used in this analysis to test that the vetoes against external sources of noise that were implemented do not 
suppress any potentially real GW events (see Section \ref{sec:vetoes}).
The hardware injection periods were not considered for the GW burst search.
Taking into account the application of the data quality criteria and the hardware injection periods, the duty cycle is finally reduced from 66\% to 55\%.
This corresponds to a total duration of 2.51 days.

\section{Description of the burst search pipeline}
\label{sec:section3}
\subsection{Overview}

The GW burst pipeline that has been used in this search is composed of several parts that allow us to select and 
analyze all segments of Virgo data and eliminate, as much as possible, the artifacts in the detector output that could mimic
a GW event.
The core of the pipeline is the GW detection statistic, described in Section \ref{sec:pipeline}, that is applied 
on pre-selected segments of data of good quality. 
This results in lists of triggers. 
As we are dealing with non Gaussian and non stationary data, many of these triggers are due to instrumental effects.
The next step consists of identifying all the sources of noise disturbances that pollute the detector output, and then define 
vetoes to be applied a posteriori on the trigger lists.
This step in this pipeline is playing an important role as we cannot suppress these artifacts by requesting coincidence 
with another detector. 
We provide details on how the vetoes have been defined and which sources of noise they help in suppressing.

\subsection{Data selection}
Instrumental problems and environmental conditions have been identified  
that temporarily affect the detector sensitivity for a GW burst search.
In a pre-selection step, we discard the periods where the detector operation is not optimal.
To do so, we have defined a list of data quality flags (DQ) that highlight the intervals of malfunctioning.
Among the DQ flags, some pertain to the saturation of the photodiodes and/or actuator signals that are used 
to keep the interferometer locked to its operating point.
The analog electronics of the control loop that maintains the laser
frequency noise below the requirements was sometimes also saturating. 
That then induces incorrect control signals and consequently a misbehaving GW strain amplitude.
An excess of events was found when the GW strain $h(t)$ reconstruction process was facing a problem, for instance
when the calibration lines were buried in the noise. 
When this happens the $h(t)$ reconstruction program sets a DQ flag stored in the data stream.
Sometimes during C7 a few frames or channels were missing due to some failure of the data acquisition system. 
That produces a hole in the continuous time series. 
The burst pipelines manage properly with the presence of holes in a segment, but to accurately compute 
the effective duty cycle we have 
defined a DQ flag pertaining to this issue.
Finally, it has been found that when an aircraft is flying above the interferometer at a relatively low altitude 
acoustic and/or seismic noise couples into the dark fringe beam and induces a strong effect in $h(t)$ that
is detected by the burst pipelines. The band limited RMS of some acoustic probes (located in the end arm buildings) 
were monitored to detect the airplanes.
All these DQ segments have been combined into a single list, taking into account possible overlaps.
The dead time induced by the application of these DQ flags on triggers list was only 0.8\%, of the total duration 
which is acceptably low.

\subsection{Event trigger generation: Exponential Gaussian Correlator}
\label{sec:pipeline}

Different pipelines have been developed to search for GW bursts. 
Many of them \cite{ref:EP, ref:WB, ref:PF, ref:Q} rely on a common principle i.e., the detection of
clusters of excess energy in a time-frequency representation of the data. 
The pipeline we use in this analysis follows the same idea;
we refer to it as the Exponential Gaussian Correlator
(EGC) as it been described in \cite{ref:EGC}, but here we give here a short summary.

The time-frequency plane can be tiled using a lattice of sine Gaussian waveforms i.e.,
\begin{equation}
\Phi(t)=\exp\left(- \frac{1}{2} \left(\frac{t}{\tau_0} \right)^2 \right) e^{2 \pi i f_0 t},
\end{equation}
where the values of the central frequency $f_0$ and typical duration $\tau_0$
can be chosen in such way as to optimize the coverage of the plane.
The idea is to consider the above waveforms as typical burst transients and to search 
for them using a matched filtering technique. 
For this reason, we designate $\Phi(t)$ as a template waveform. 
The EGC computes the cross-correlation of the data with the templates, namely
\begin{equation}
C(t) = \frac{1}{N} \int_{-\infty}^{+\infty} \frac{\tilde{x}(f) \tilde{\Phi}^*(f)}{S(f)} e^{2\pi i ft} df,
\end{equation}
where $\tilde{x}(f)$ and $\tilde{\Phi}(f)$ are the Fourier transforms of the data and 
template, and $S(f)$ is the two-sided power spectral density of the noise.
$N$ is the normalization factor of the templates defined as 
\begin{equation}
N=\sqrt{\int_{-\infty}^{+\infty} \frac{|\Phi(f)|^2}{S(f)} df},
\end{equation}
 
For the present analysis, the power spectral density has been estimated over data segments of 600 second 
duration.
To define the template lattice, the parameterization using the
quality factor $Q_0= 2\pi \tau_0 f_0$ instead of the duration $\tau_0$
is preferred. 
Using the algorithm of \cite{ref:arnaud03}, we tile the parameter space
$(f_0, Q_0)$ with a minimal match of 99\%. The parameter ranges 150 Hz
$\leq f_0 \leq$ 2 kHz and $2 \leq Q_0 \leq 16$ correspond to the
frequency band of best sensitivity for Virgo during the C7 run. 
This generates 420 templates; 
the shortest and longest ones have 0.31 ms and 34 ms duration.

The quantity $\rho=\sqrt{2 \times |C|^2}$ is the signal-to-noise ratio (SNR),
which we use as a detection statistic. 
It depends on the analysis time $t$, the template frequency $f_0$, and quality factor $Q_0$. 
It thus defines a three-dimensional representation map of the data \cite{ref:EGC}
in which we search for a local excess as compared to typical noise fluctuations.

We first apply a low threshold (SNR of 3.3) to the map.  
We form clusters with the surviving pixels of energy $|C_i|^2$ with a two pass procedure. 
The pixels are grouped into clusters using the Hoshen-Kopelman algorithm \cite{ref:HK}. 
Once the cluster is formed, all pixels that have a SNR lower than 5 are removed.
Then, all clusters overlapping in time or separated by less than 50 ms are grouped 
together once again.  
We produce the final list of triggers by requesting that the cluster SNR 
($\sqrt{2 \times \sum{|C_i|^2}}$) is larger than 11.3.
This choice for the threshold is a good compromise between not having too many triggers
and not losing too much detection efficiency.

The triggers' information is extracted from the clusters. The peak
time, peak frequency and the trigger SNR are defined by the pixel that
has the highest SNR. The duration and the frequency bandwidth of the
trigger take into account all the pixels in the cluster. 
The timing resolution of EGC has been estimated for different kinds of waveforms
\cite{ref:EGC, ref:LV_Burst}. 
For most signals used in the benchmark test, the timing resolution of EGC is smaller 
than 1 ms for SNR of 10, but it can be as large as a few ms.

\subsection{Vetoes}
\label{sec:DQ}
The search for GW burst events presented here was carried out on
real data that are non-Gaussian and non-stationary. This
prevents the use of theoretical estimates of the expected false
alarm rate. On the contrary, the background rate must be extracted
from the data itself. 
In the multi-detector context, the background can be estimated by
computing the rate of (purely accidental) coincidences between trigger
lists that were time shifted with random delays \cite{ref:LIGO_burst_S2, ref:LIGO_burst_S4}. 
We cannot resort to such procedure in the case of this study since we only have
the Virgo C7 data set at our disposal.
There is a need to initially identify, in an thorough way, all the
sources of noise that generate large SNR triggers that can mimic GW
burst events; these types of noise events have to be suppressed before we try to estimate the background event number.

We define a methodology that guaranties that we do not discard any true
GW event as long as they do not coincide with a bad quality data
period.
Initially, there was the study of all the sources of noise that
generate large SNR triggers, or more generally an excess of triggers,
and the construction of vetoes to suppress these triggers. The study
was performed on a subset of the C7 dataset (the playground segment)
chosen arbitrarily before starting the GW burst search, in order to
avoid bias. Our main concern in the veto development process was to
minimize the dead time introduced by the data quality selection, while
assuring a high level of noise trigger rejection.
   To perform the background studies and develop the veto strategy we
used the longest C7 science segment; this segment was 14 hours in
duration (22\% of the whole dataset used for the GW burst search). 

Several transient detection algorithms have been applied to the playground
dataset to identify and characterize the sources of noise
that generate the high SNR triggers that mimic a GW burst event.
Mean Filter searches for an excess in a moving average computed 
on whitened data \cite{ref:MF}. 
Peak Correlator is a matched filter using Gaussian waveform templates \cite{ref:MF}. 
The Wavelet Detection Filter is based on a multi-resolution 
time-frequency transform applied on whitened data \cite{ref:C7_GRB}.
Results of these pipelines have been compared to
ensure that all sources of transient events have been discovered.
These detection algorithms search for a short duration excess of
energy, but using different methods. All of the trigger lists show a
rather large excess of events compared to expectation from Gaussian
noise; there was a definite excess of large SNR triggers. 

Two categories of large triggers have been identified; the first one
corresponds to short glitches correlated to a glitch also present in
an auxiliary channel. These auxiliary channels includes the
available environmental monitoring and control loop signals (see Section \ref{sec:glitches}).
The second category of glitches have been found to be due to the 
so-called "bursts of bursts" that we describe in Section \ref{sec:loudest}.
Specific vetoes against these two sorts of glitches have been developed as explained in Section \ref{sec:vetoes}

\subsubsection{Origin of the short duration glitches}
\label{sec:glitches}
To veto all periods of data during which instrumental or environmental problems occurred 
and generated a glitch in the GW channel data, one first needs to figure out which auxiliary channels
are the most sensitive to these particular sources of noise. 
The by-eye scanning of the loudest events observed in the playground dataset already gave hints to 
potentially useful auxiliary channels.
However we subsequently performed a systematic analysis using many 
channels recorded by Virgo and using the Mean Filter pipeline. 

We have found that three channels were particularly interesting for vetoing glitches in the GW strain 
amplitude channel $h(t)$.
It has been seen that many high SNR burst events are coincident in time with the realignment of the 
quadrant photodiodes by some stepping motors. 
These quadrant photodiodes, used for the interferometer automatic alignment system, are located at the 
output of the interferometer, on the same optical bench hosting the photodiodes used to detect the 
interferometer dark port signal. 
The stepping motors were generating very loud acoustic and mechanical noise that coupled into the dark fringe 
beam nearby. Each time a quadrant photodiode is moved there is a large increase in the signal measured by 
an accelerometer probe located on the detection bench. 
When the quadrant photodiodes were not re-aligned (horizontal and vertical positions remain constant)
the RMS of the accelerometer probe signal remained low. 
The rate of the quadrant photodiodes' realignment was not constant over the run;
initially the alignment occurred every 3 seconds, but this was then reduced to every 300 seconds at the end 
the C7 run after the discovery of the problem (the parameters of the process responsible for the quadrant 
photodiodes realignment have subsequently been changed to diminish this problem). 
This acoustic/mechanical noise generated glitches in the GW strain amplitude channel whose frequency content was around
550 Hz. Due to the fact that the excess of acoustic noise could last up to 1 second, which is long for a 
glitch finding algorithm, we used the RMS computed over 100 ms of data from an accelerometer located on
the detection bench instead of the Mean Filter triggers to create a veto (called 'Seismic' in the writing below). 
This allowed a more precise definition of the GPS time of the maximum of the excess of noise. 
This source of glitches has been found to be dominant in the playground dataset, but other (and rarer) 
types of glitches have also been found in the C7 data. 
Indeed, events in the Second Stage Frequency Stabilization System correction signal were observed in
coincidence with $h(t)$ glitches. 
Some remaining loud glitches discovered in the GW strain amplitude channel have been found to be coincident in
time with a dip in the power of the beam reflected by the interferometer toward the laser as recorded by the 
so-called B2 photodiode.
The light impinging on the B2 photodiode is sensitive to the power recycling cavity length change. 
The origin of the power dips have not been understood, but the effect on the GW strain amplitude
was demonstrated and the safety of this veto using an optical auxiliary channel has been carefully studied, 
and is discussed below. To define an event by event veto for the two latest categories of glitches we used the 
Mean Filter triggers. These vetoes are hereafter referred to as 'SSFS' and 'B2'.

\subsubsection{Bursts of bursts}
\label{sec:loudest}

\begin{figure}[h]
\begin{center}
\epsfig{file=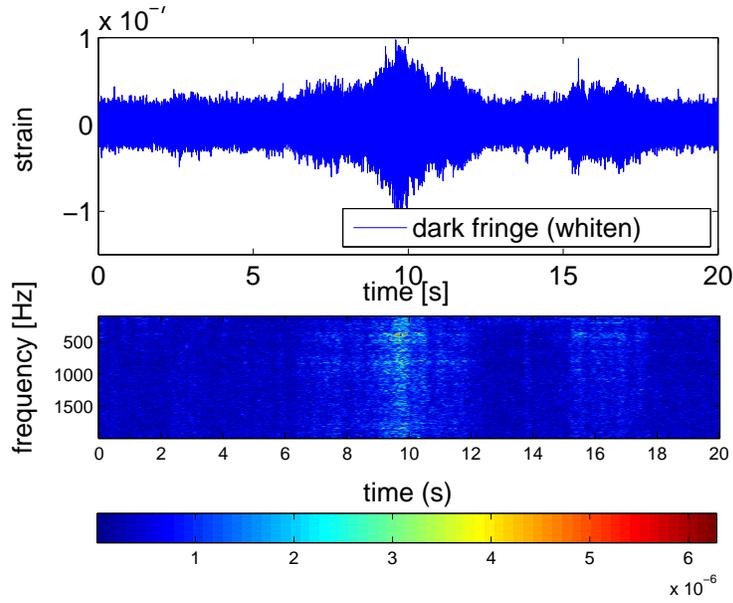, height=8cm}
\caption{\label{fig:a_bob} An example of a Burst of Bursts (BoB) around t $\simeq$ 10 s. The upper plot shows 
the whitened interferometer dark port channel amplitude as a function of time, 
while the bottom plot is the corresponding spectrogram.}
\end{center}
\end{figure}

We noticed that the burst triggers were not uniformly distributed over time but seemed to show up in bunches 
lasting up to a few seconds. 
These events have been called Bursts of Bursts (BoB). 
FIG. \ref{fig:a_bob} shows the whitened GW strain channel's amplitude and its spectrogram during a data 
period that contains a burst of bursts. 

\begin{figure}[h]
\begin{center}
\begin{tabular}{cc}
\epsfig{file=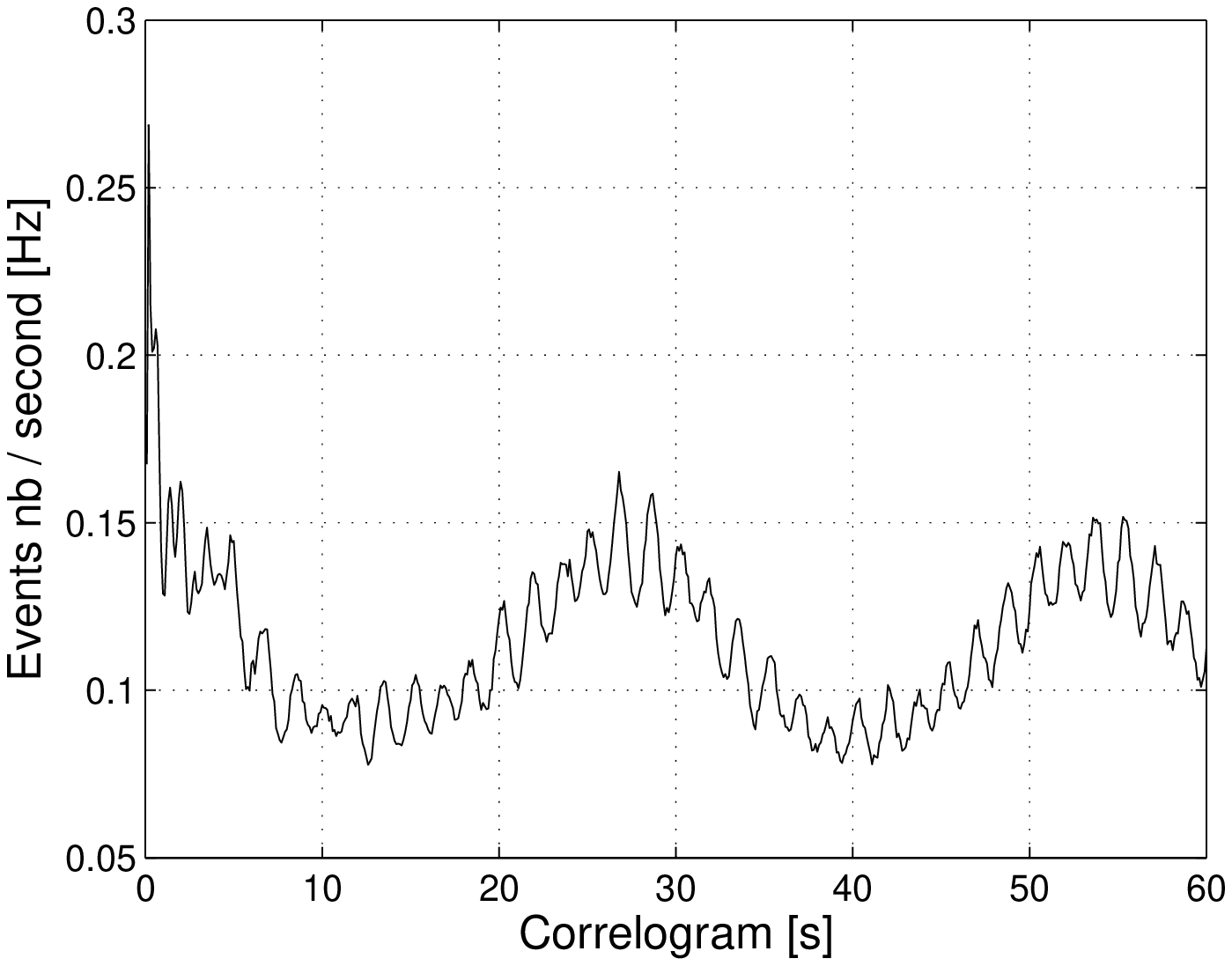, width=9cm} &
\epsfig{file=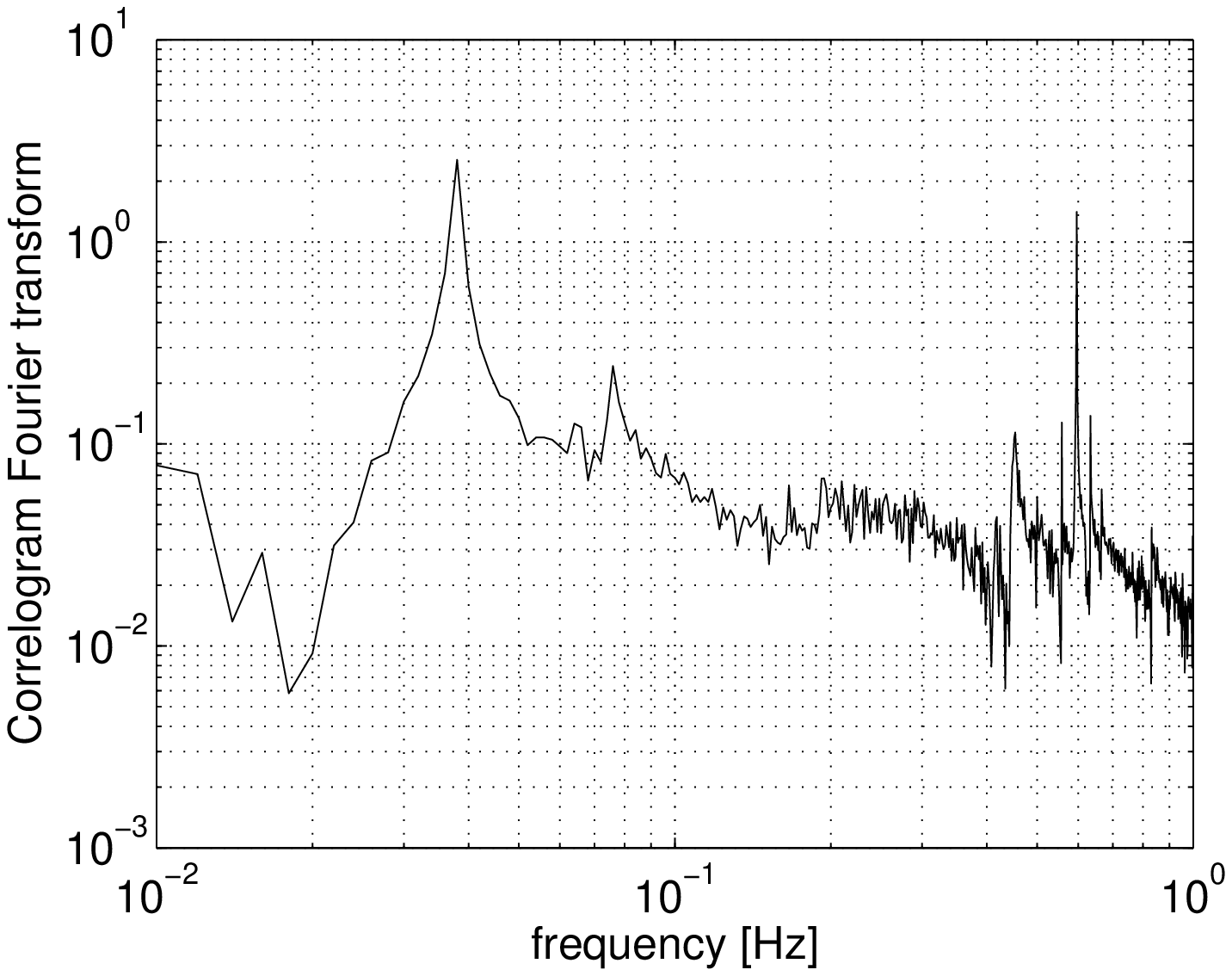, width=9cm} \\
\end{tabular}

\caption{\label{fig:correlogram_27s} Left: the correlogram of the triggers found by the EGC pipeline in the C7 dataset.
A peak around T $\simeq$ 27\,s is present, with a replica at 54\,s. A 0.6 Hz modulation is also visible, and this 
frequency corresponds to the pendulum mode of the last stage of the Virgo suspension. Right: the Fourier transform of 
the correlogram. The 37 mHz mechanical resonance between the two last stages of the Virgo suspension is 
clearly visible.}
\end{center}
\end{figure}

The spectrogram shows that the event has a broadband frequency content contrary to 
most of the events caused by external disturbances.
This broadband spectral signature provided clues that helped to identify the 
BoBs as local increases of the noise 
level due to an increase of the coupling factor between the frequency noise (which is only a factor 2 
lower than the shot noise at high frequency for 
C7) and the interferometer's dark port channel. 
The presence of a peak around T $\simeq$ 27 s in the correlogram of the burst events (see FIG. \ref{fig:correlogram_27s})
provided evidence that the residual angular motion of the mirrors could be playing a role. 
Indeed, 27 s corresponds to the period of a mechanical resonance between the two last stages of the Virgo suspension
that, if excited, may induce mirror angular degrees of freedom motion \cite{ref:suspension}.

It turns out that the mirrors' alignment control system was not optimally working during C7 and the angular tilts 
of the mirrors induced an increase of the coupling of the laser frequency noise with the dark port amplitude.
The BoB events represent a rather large fraction of the burst triggers in C7, as can be seen 
in FIG. \ref{fig:correlogram_27s}, where the number of triggers in the correlogram fluctuates
by up to $\sim$ 50\%. This is also visible by the height and width of the 37 mHz peak in the Fourier 
transform of the correlogram. 
The BoBs contribute partly to the highest SNR triggers, but mainly to the moderate SNR values.


These non-stationary noise periods must be eliminated from the analysis since they generate an excess of 
noise triggers and they 
correspond to periods where the Virgo sensitivity is degraded compared to the norm. 
To identify those periods we used the band limited RMS of the dark port channel amplitude around 1111Hz, which is the 
frequency of the line injected to measure the laser frequency noise component in the dark port channel.
This line allows one to measure the common mode noise level in the GW strain amplitude, 
and especially the laser frequency noise. 
Actually, measuring the height of the line allows one to monitor the variation of the 
common mode coupling factor \cite{ref:gouaty}.
We computed the band limited RMS of the dark port channel amplitude around 1111Hz over 10 Hz (1106-1116 Hz), 
at a sampling rate of 20Hz.
The high values of the band limited RMS correspond to periods where the non-stationary excess of noise is large.
That allowed us to define the '1111Hz' veto.\\

\subsubsection{Veto parameters tuning}
\label{sec:vetoes}

Once the channel and filter have been identified, one needs to tune the parameters of the veto, 
namely the threshold and the duration of the vetoed window around the time of the glitch. 
The parameters are adjusted to maximize the veto significance, keeping
the dead time (i.e., percentage of vetoed science time) below a reasonable value. 
The significance is defined as the ratio of the number of triggers (presumed to be
instrumental glitches) that the procedure vetoes by the square root of the number of triggers it would accidentally veto 
if there is no physical link between the glitches of the auxiliary channel and $h(t)$ \footnote{The significance = $\frac{N-n}{\sqrt{n}}$ 
where $N$ is the number of coincidences and $n$ is the estimation of accidental events assuming Poisson statistics.}. 
The significance measures the excess of coincident events in standard deviation units.
We used the GW strain amplitude triggers of the EGC pipeline to tune the vetoes as we wanted an optimal background
rejection for this analysis.
The rate of accidental coincidence was measured by artificially time shifting the list of auxiliary 
glitches while the number of vetoed triggers is obtained for the zero-lag.
We vary the threshold on the auxiliary channel output (SNR or RMS) and the duration of the veto window.
For each value we compute
the dead time, the efficiency (fraction of EGC triggers that are vetoed) and the use percentage 
of the veto (fraction of auxiliary channel triggers that veto an EGC trigger).
A good veto must have a high use percentage \cite{ref:vetoes}.
The size of the veto window must be larger than the peak time difference in order to conservatively veto the region 
around the glitch, and can be as large as the total duration of the auxiliary channel trigger. 
Values from 100 ms up to 600 ms have been tested.

\begin{figure}[h]
\begin{center}
\epsfig{file=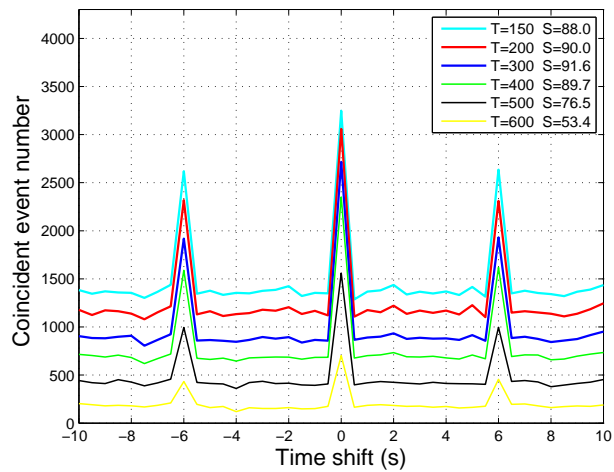, width=9cm}
\caption{\label{fig:SEDBsig} Number of coincident triggers between the GW strain amplitude and the auxiliary 
channel that has been used to define the Seismic veto, as a function of the time shift introduced in the Seismic 
veto list. The different curves have been obtained for different threshold values (T). S is the significance.
The peaks correspond to the stepping motor that have been activated each 6 seconds during the playground dataset. 
The different curves have been obtained for various values of the threshold applied on the auxiliary channel RMS.
The maximal value of the significance is obtained for a threshold of T=300.}
\end{center}
\end{figure}

FIG. \ref{fig:SEDBsig} shows the number of coincident triggers between the GW strain amplitude and the auxiliary 
channel that has been used to define the Seismic veto, as a function of the time shift introduced in one of the lists. 
In this case, the auxiliary channel triggers are the RMS of the channel computed at 10 Hz.
Different values of the threshold have been tested (from 150 up to 650 by steps of 50).
The maximal value of the significance is obtained for a threshold of 300. 
The threshold is confirmed by the fact that below this value the veto efficiency reaches a plateau.
The choice of the duration of the veto window is based on a compromise between a high use percentage and a low dead time
for the value of the threshold chosen.
We checked that the optimal value, 500 ms, matches the width of the time difference between the auxiliary channel RMS 
and the EGC triggers.

The same procedure has been applied to tune the parameters of the 'SSFS' and 'B2' vetoes 
using the playground dataset.
Some differences, compared to the Seismic case should be noted.
For these two vetoes, built on the triggers generated by the Mean Filter glitch finding algorithm, 
we used the information of the trigger duration to define the veto duration, in addition to a fixed minimal 
window size. 
For the 1111Hz veto, the duration of the window was defined by the time when the band RMS was above the threshold.
The veto duration was then symmetrically enlarged.
Given the nature of the BoB events (high EGC trigger multiplicity and rather noisy periods) we could not use the significance
to optimize the threshold as it leads to setting the threshold at a very low value corresponding to an unacceptable large 
dead time. Instead, we set the threshold such that the dead time remained below 20\%.
We checked that this choice assures that the veto was able to suppress a large fraction of the
BoB events.
The veto definitions and the chosen parameters are reported in TABLE \ref{tab:veto_tuning}.

\begin{table}[h]
\begin{center}
\begin{tabular}{c|c|c|c|c}
\hline
Veto    & Auxiliary channel quantity & threshold tuning method & threshold & time window \\
\hline
1111Hz  & Band RMS (computed at 20Hz)&  dead time          & 12   & period above the threshold +/-300 ms \\
Seismic    & RMS (computed at 10Hz)     &  significance    & 300  & 500 ms (fixed)                  \\
SSFS    & SNR of the Mean Filter triggers     &  significance    &  6   & Mean Filter trigger duration - minimal value: 300ms \\
B2    & SNR of the Mean Filter triggers     &  significance    & 10   & Mean Filter trigger duration - minimal value: 300ms \\
\hline
\end{tabular}
\end{center}
\caption{\label{tab:veto_tuning} Definition and parameters of the four event by event vetoes that have been developed and 
tuned for this analysis.}

\end{table}

\subsubsection{Veto safety}

The safety of a veto is of fundamental importance as we do not want to inadvertently eliminate any real GW event.
Vetoes based and dedicated to suppress coincident glitches in the GW strain amplitude channel must be safe
with respect to real GW events. 
The DQ flags and vetoes which suppress deleterious periods of data can be potentially unsafe 
since the source of noise is independent from the effect of a real GW impinging upon the detector.
On the contrary, we must be sure that a veto based on an auxiliary channel that has some glitches remains silent 
when a real GW event is passing through the detector. 
More precisely, environmental channels such as acoustic or seismic probes are expected to be safe, whereas
vetoes constructed on optical or interferometer control channels may be unsafe, 
since a GW event will generate a change in photodiode signals that 
can be used within the feedback loop of some control systems. 
To test the safety of an auxiliary channel, one can examine the periods where hardware injections of 
fictitious GW signals are inserted into the interferometer.
A deterministic force applied on the North Input (NI) mirror, induces a variation of the length of the North arm 
Fabry-Perot cavity which then mimics the effect of a real GW on the interferometer.
During C7 there were two periods of hardware injections, and these included two different type of burst waveforms: 
60 Sine Gaussian ($f$=920 Hz, $Q$=15, and $f$=460 Hz, $Q$=15), and 33 Gaussian ($\sigma$=1ms).
Both of these signals were injected with a SNR of 15 using 
a sensitivity curve taken just before the run started (the real SNR of the hardware injections is somehow 
different due to the sensitivity variation during the run). 
The critical point concerns the safety of the SSFS and B2 vetoes, which could in principle be unsafe. 
None of the hardware injection signals have been vetoed by the Seismic, SSFS and B2 vetoes.
This establishes that our vetoes were safe.
The anti-BoB veto 1111Hz which was built using directly the gravitational wave channel was, by construction, unsafe for GW 
burst event whose frequency content is around 1111 $\pm$ 5 Hz.
We could not check that the veto was unsafe for these signals using the hardware injections as no signal 
with enough amplitude in this frequency range has been injected, but we should consider that the GW burst analysis presented 
here is insensitive for GW signals in the 1106-1116 Hz frequency band.
Besides, the safety of a veto assumes that its dead time remains small. 
This is the case for all the vetoes used in this analysis except the 1111Hz veto whose large dead time (16.1\%) might
suppress a real GW burst event.
The anti-BoB veto 1111Hz suppressed 2 hardware injections, which is less than what we could 
foresee given the large dead time of this veto.

\section{Output of the GW burst search pipeline}
\label{sec:background}

\subsection{First results}

\begin{figure}[h]
\begin{center}
\epsfig{file=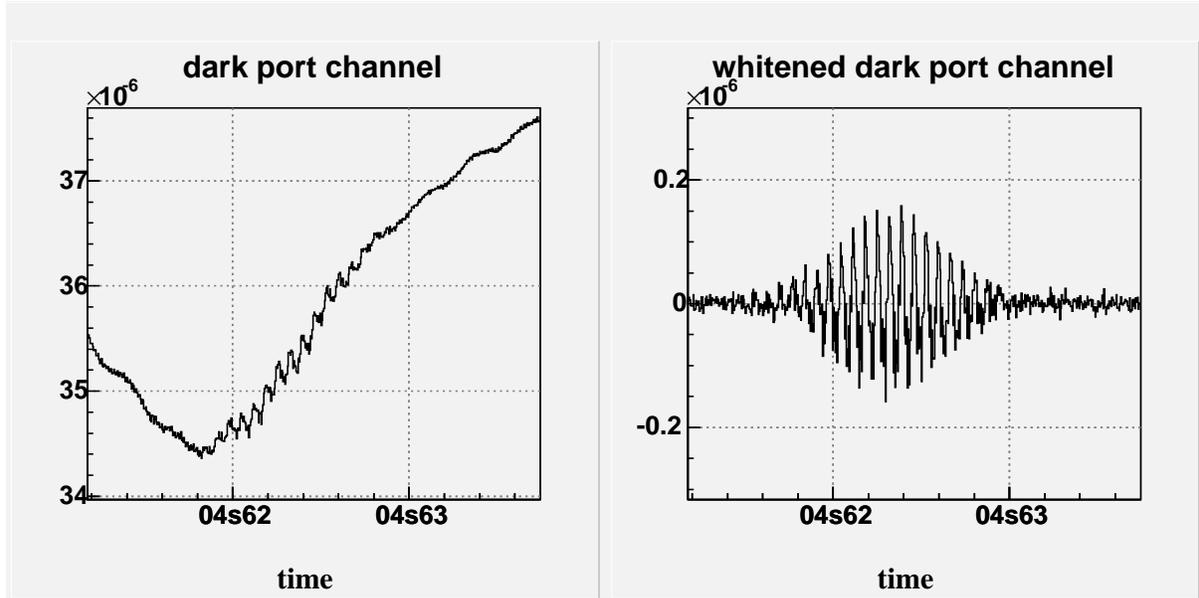, height=8cm}
\caption{\label{fig:C7_EGC_SG} Time series in the dark port channel (left) and the whitened dark port channel (right) 
of the loudest event detected by the EGC pipeline in the search for GW bursts in the C7 data. This event was seen with 
a SNR of 69.7 at a frequency about 1466 Hz. }
\end{center}
\end{figure}

We have generated the veto lists for the full C7 data set applying the tunings that have been 
defined using the playground segment as explained in Section \ref{sec:vetoes}. 
After applying the vetoes on the whole dataset, the remaining largest SNR triggers were 
studied in order to assess their compatibility with a GW burst signal. 
At this stage, we ended up with a single rather large SNR event that has been subsequently thoroughly studied. 
This very peculiar event has been observed outside the playground segment.
Only one event of this type has been found in the full run, at least with such a SNR.
It looks like a Sine Gaussian signal, as shown in FIG. \ref{fig:C7_EGC_SG},
with a SNR of 69.7, for one of the Sine Gaussian templates (f=1466Hz, Q=15.5, $\sigma=Q/\Pi f$=3.4 ms);
All environmental channels have been checked around this period using different glitch finding algorithms.
Nothing suspect was found in any of the channels.
However, the event was visible in the other phase of the demodulated dark port signal.
We concluded that this event could not have been generated by a GW burst, but more likely could be due
to an experimental problem, not occurring in the chosen playground segment. 
Indeed, the modulation phase angle was tuned such that the effect of a GW crossing the interferometer was contained in one 
demodulated phase (ACp) of the dark port signal, while the other (ACq) should not be perturbed.
This observation lead us to develop an event by event veto (called 'PQ') based on the ratio of the SNR of time coincident 
triggers in the two demodulated phase signals as proposed in the literature \cite{ref:PQMon, ref:Hanna,ref:Christensen05}.
Despite the fact that such kind of events have not been found in the playground segment as a significant source of
loud noise events, we decided to develop this veto and to apply it on the full data set.
The other loudest high SNR events have been found compatible with BoB-like events.

\subsection{Necessity of an a posteriori veto}

\label{sec:PQ}

The in-phase signal contains {\it a priori} the GW strain amplitude, provided that the demodulation phase is well tuned.
An error in the demodulation phase induces a small coupling of the GW signal with the quadrature channel.
However, the ratio of the GW energy seen in the two phases is expected to remain high.
The ratio is expected to be proportional to $\frac{1}{sin(\delta \phi)}$, where $\delta \phi$ is the error on the 
demodulation phase \cite{ref:Hanna}. Unfortunately we do not know precisely the value of $\delta \phi$ during C7.
A real GW event will be seen with a SNR in ACp much higher than in ACq.
On the contrary, ACq will be sensitive to glitches in signals which are related to common mode noise.
It may happen that some source of noise can affect both quadrature signals with similar strength.
This is especially the case for a dust particle crossing the laser beam before the output dark port 
photodiode where the beam is especially small in width.
A key point of such a veto is to assure that no real GW events would be suppressed, and therefore one
would need to develop a veto with a rather good security factor.
To do so, we used the hardware injections to verify safety and develop the characteristics of this veto.

\begin{figure}[t]
\begin{center}
\epsfig{file=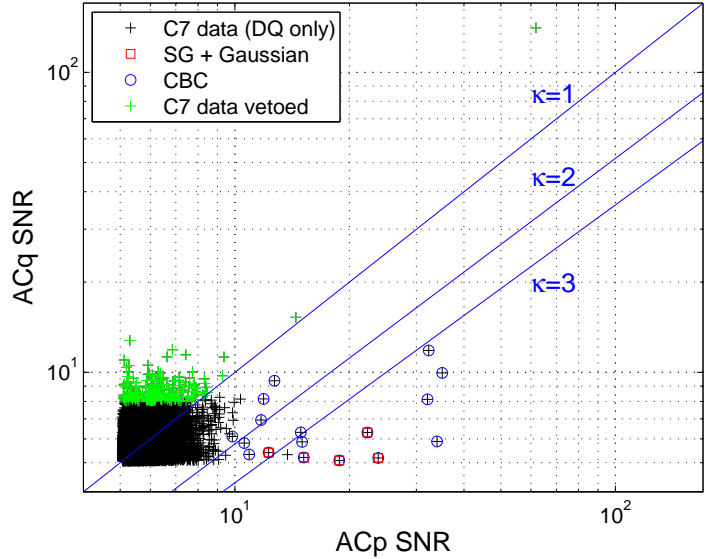, height=8cm} 
\caption{\label{fig:C7_PQ_EGC} SNR of the coincident EGC triggers in the two quadrature demodulated channels of the output 
dark port signal: ACp (in phase) and ACq (in quadrature). The time coincidence window of the triggers seen in the two channels is
10 ms. The Seismic, SSFS, B2 and 1111Hz vetoes have already been applied. 
The peculiar event appears very isolated with respect to the other triggers with its ACq SNR of 141. 
The hardware injection signals, which were seen in both of the two demodulation phase channels, are indicated by the circles
for burst-like signals and by the squares for the compact binary coalescence (CBC) signals.}
\end{center}
\end{figure}

EGC triggers have been generated for the dark port ACp and ACq channels and all events with SNR above 5 were considered.
We then required time coincidence between the two sets of triggers using a window of $\pm$10 ms. 
No coincidence in frequency has been required.
The hardware injection signals were detected in the ACp triggers with an efficiency of 98\%.
Unfortunately during C7, the injected SNR was not, on average, very large and consequently
the signal energy in the ACq
channel could not be expected to be very high (below the pipeline threshold at SNR=5).
Among the 91 detected hardware injections only 5 had a coincident trigger in the ACq phase channel 
within 10 ms; this tends to indicate that a large fraction of the GW energy would be contained in the ACp channel 
as expected; this is displayed in FIG. \ref{fig:C7_PQ_EGC}.
The ratio between the SNR detected in the two demodulation phase channels in coincidence is
\begin{equation}
\kappa = \frac{SNR_{ACp}}{SNR_{ACq}} \nonumber
\end{equation}
 
The very low event statistics (5 events detected simultaneously in ACp and ACq) 
encourages us to stay conservative when defining the PQ veto parameters.
We checked that the use of a $\pm$10 ms window limits the accidental coincidence event rate to 0.2 over the period of the 
hardware injections (assuming a Poisson trigger rate).
This excludes the fact that we have more than one accidental association with an ACq trigger for the fix burst 
hardware injections as shown in 
FIG. \ref{fig:C7_PQ_EGC}.
In this figure, the strange and unique high SNR event is totally isolated from the rest of the events. 
This event has $\kappa \simeq 0.42$ and a SNR in the ACq channel of 141; these facts totally exclude the possibility that 
this event has been generated by a real GW.

The goal of the PQ veto is to suppress high SNR events in the ACq channel that induce a 
transient in the dark port channel amplitude.
That is why it is reasonable to consider only large SNR ACq triggers.
A threshold is applied on the SNR of the ACq triggers ($SNR_{ACq}$). 
The other veto parameter threshold pertains to the ratio $\kappa$.
We did not try to optimize these two parameters, 
but rather we keep a conservative attitude with respect to the hardware injection signals' position in the two-dimensional plane 
that shows the SNR of both quadratures for the coincident triggers (see FIG. \ref{fig:C7_PQ_EGC}).
The veto list segments have been defined as the periods during which $SNR_{ACq}>8$ and $\kappa <1$.
All burst hardware injection signals pass these conditions with a safety factor since they are all below 
the $\kappa=2$ line.
The starting and ending times of the veto segments are given by the ACq trigger times.
The dead time of the PQ veto is 0.07\%, which is very acceptable.

\subsection{Final results}
 
TABLE \ref{tab:veto_deadtime} gives the dead time of the five event-by-event vetoes which have been used in this analysis.
The total dead time amounts to 21.2\% taking into account overlaping veto segments.
This is a rather high value, mainly because of the presence of the non-stationary excess of noise due the looseness of the
mirror's angular degree of freedom control during C7.

\begin{table}[h]
\begin{center}
\begin{tabular}{c||c|c|c|c|c||c}
\hline
Veto name   &     Seismic        & SSFS    & B2      & PQ      & 1111Hz   & all vetoes\\
\hline					    
dead time   &  3.75\%       &  2.6\%   &  0.03\%  & 0.07\%   & 16.1\%   &  21.2\%   \\
\hline
\end{tabular}
\end{center}
\caption{\label{tab:veto_deadtime} Dead time of the different vetoes that have been found useful. The values are computed 
for the full C7 dataset.}
\end{table}

\label{sec:results}
\begin{figure}[t]
\begin{center}
\epsfig{file=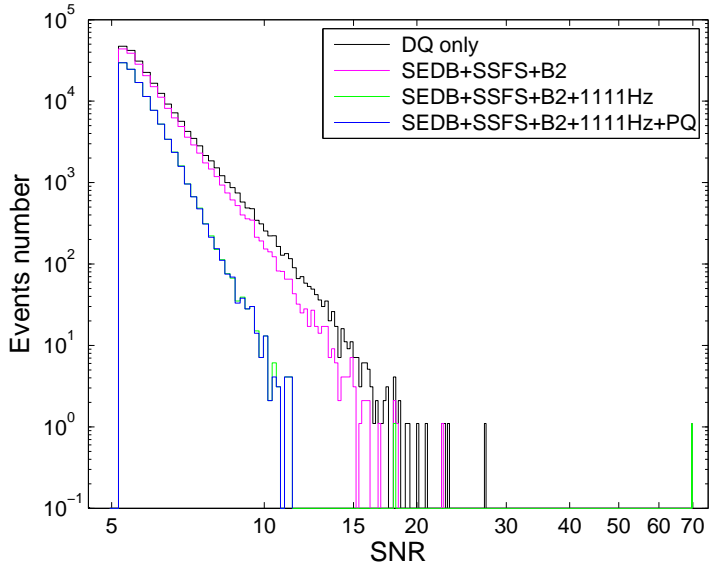, height=8cm}
\caption{\label{fig:C7_background_final} SNR distribution of the triggers obtained on the GW strain amplitude channel by 
EGC after applying all vetoes on the full C7 data set. }
\end{center}
\end{figure}

\begin{figure}[t]
\begin{center}
\epsfig{file=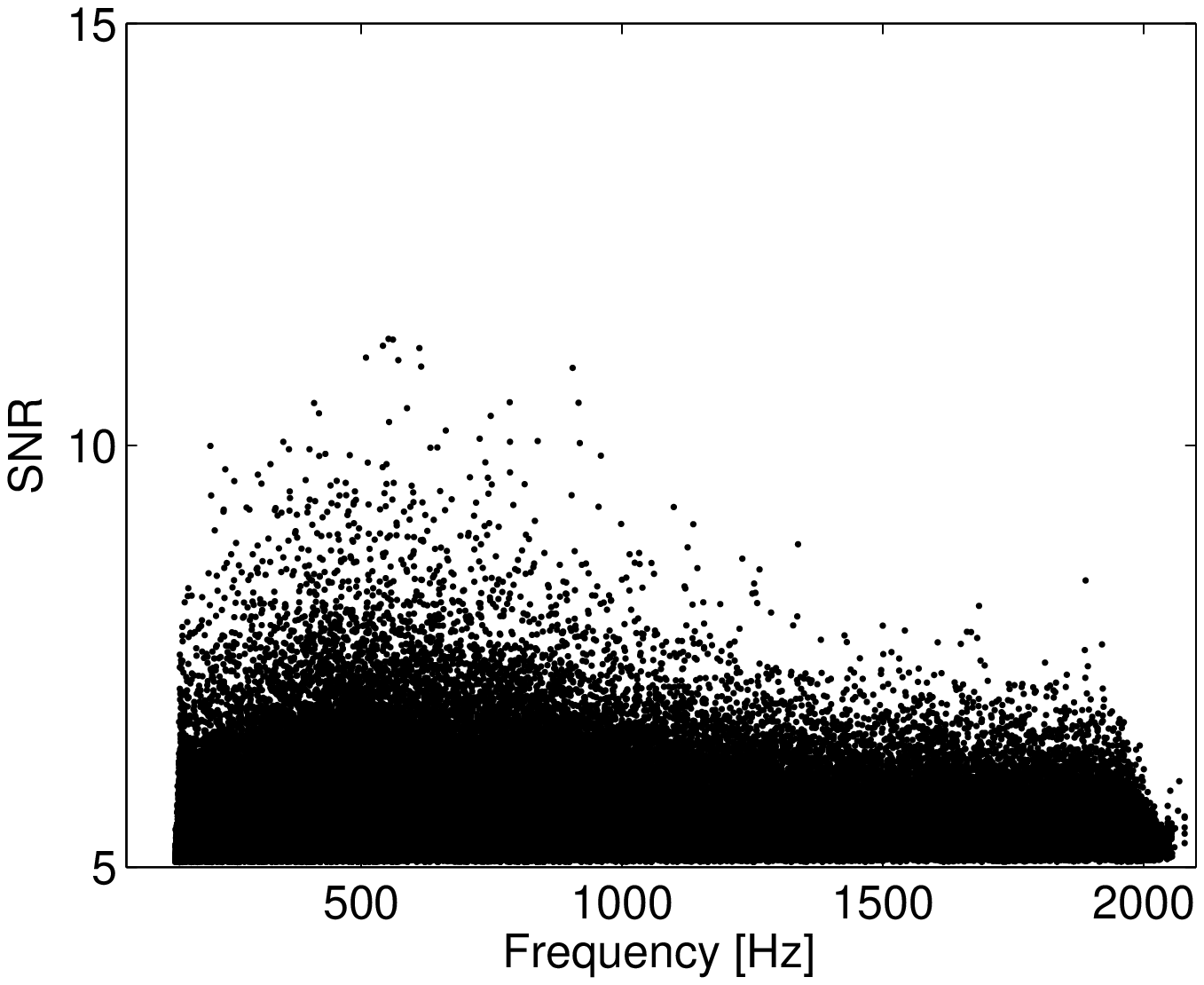, height=8cm}
\caption{\label{fig:C7_frequency_final} SNR as a function of the frequency of the triggers obtained on the GW strain amplitude 
channel by EGC after applying all vetoes on the full C7 data set. 
The distribution around 2 kHz is due to the templates' placement; 
some Sine Gaussian templates have a central frequency which can slightly exceed 2kHz.
}
\end{center}
\end{figure}

FIG. \ref{fig:C7_background_final} shows the triggers' SNR distribution obtained 
from the full C7 data set, after having applied consecutively the five vetoes (note that the highest SNR event is 
off of the scale of the plot).
The four anti-glitch vetoes (Seismic, SSFS, B2 and PQ) suppress 10.6\% of the total number of triggers and play an important role 
for the large SNR ($>$10) events since 51.3\% of them are eliminated.
The anti-BoB veto is critically important, and its efficiency on EGC triggers is impressive;
45.6\% of the remaining triggers are eliminated, and 97.6\% of the remaining loudest ones (SNR $>$ 10). 
FIG. \ref{fig:C7_frequency_final} shows the SNR of the remaining triggers as a function of their frequency.
The application of all these vetoes allows us to reduce, by a factor 2, the number of events for all EGC triggers above SNR=5.
98.8\% of the EGC events above SNR$>$10 were eliminated.
21 triggers above SNR=10 remain at the end. 
The loudest has a SNR of 11.26, and although it was not vetoed, 
it seems to be due to a short period non stationary Virgo noise increase.

\section{GW burst search analysis sensitivity}
\label{sec:section5}
\subsection{Generation of the injected waveforms}

\label{sec:efficiency}

The GW burst search presented in this paper is intended to be efficient for all 
waveforms as long as they have energy 
concentrated in time and  frequency in the range 150-2000 Hz.
However, since the Virgo sensitivity is not constant over this frequency band, 
the detection efficiency for a given signal strength will depend on the frequency of the signal.
In order to assess the sensitivity of the detector to dissimilar kinds of
signals at different frequencies, we used simulated waveforms of diverse types. 
These were added to the Virgo GW strain amplitude (software injection) and analyzed by the EGC pipeline, 
similar to how the C7 triggers were produced.
These injected signals were randomly distributed in time during the C7 run, assuring a minimal separation of 
60 s~\cite{ref:siesta}.


The waveforms investigated in this analysis were of two types. One group consisted of ad-hoc waveforms, such as Sine 
Gaussians and Gaussians,
whose detection efficiency is expected to depend mainly on the central frequency, bandwidth and duration of the waveform.
On the other hand, to test the efficiency of this analysis to detect BBH merger and ring-down GW signals we used 
waveforms provided by numerical relativity simulations \cite{ref:gsfc}.
We also used a signal pertaining to a star core collapse, referred to as A1B2G1, from the supernova simulation catalog 
available in \cite{ref:garching}.

The amplitude of the GW associated with an astrophysical source depends on the distance of the source from the 
Earth. 
To quantify the strength of a GW burst, we used the root sum squared of the strain amplitude at the Earth 
without folding in the detector antenna pattern
\begin{equation}
\label{eq:hrss}
h_{rss} = \sqrt{\int \left( |h_\times(t)|^2 + |h_+(t)|^2 \right) dt},
\end{equation}
where $h_+$ and $h_\times$ are the two GW polarizations that we express below for the different cases relevant to our analysis.
We assume in Eq. (\ref{eq:hrss}) that the source is optimally oriented with respect to the detector.
\subsubsection{Sine Gaussian and Gaussian waveforms}
For the Sine Gaussian and Gaussian signals, the detection efficiency is expected to depend mainly on the central frequency, bandwidth and duration of the waveform. For these signals the strain sensitivity $h(t)$ of the detector can be
written as 
\begin{equation}
h(t) = F_+(\theta, \phi,\Psi) h_+ (t) + F_\times(\theta,\phi, \Psi) h_\times (t),
\label{eq:antennapattern-adhoc}
\end{equation}
where $F_\times$ and $F_+$ are the antenna response functions characterized by
the source position $(\theta, \phi)$  and polarization angle $\Psi$ relative to the detector~\cite{ref:BP}. 
Specifically, for Sine Gaussian injections we used circular polarized waveforms described by
\begin{eqnarray}
h_{\times}(t) = h_0 ~ e^{-2(\pi f_0 t / Q_0 )^2 } ~ \sin(2\pi f_0 t), \nonumber \\
h_{+}(t)      = h_0 ~ e^{-2(\pi f_0 t / Q_0 )^2 } ~ \cos (2\pi f_0 t).
\end{eqnarray}
The typical duration of a Sine Gaussian signal depends on the parameter $Q_0$ and the central frequency $f_0$ as $\frac{Q_0}{2\pi f_0}$.
Several values of $f_0$ ranging from 100 Hz up to 1797 Hz and $Q_0$=3, 8.9 and 9
were used.
Their narrow-band feature allows the testing of the pipeline performance in a given frequency region.

For Gaussian injections, we used linearly polarized waveform given by 
\begin{eqnarray}
h_+(t) &=& h_0\, e^{-t^2 / \tau^2},\\
h_\times&=& 0,
\end{eqnarray}
where $\tau$ is the width of the signal.
Gaussian waveforms are of interest in the GW burst search because many predicted signals associated with core collapse 
GW emission just after the bounce have large peak structures. 
We chose the values of $\tau$ to lie between 0.1 and 6 ms. 
In the case of linearly polarized Gaussians, the angles $\theta$, $\phi$ have been randomly chosen. 
For the circularly polarized Sine Gaussians the randomization has been done over the angles $\theta$, $\phi$ and $\Psi$.

\subsubsection{Astrophysical waveforms}
For the core collapse signals, we have chosen only one waveform produced by a core collapse 3D simulation conducted with numerical 
general relativistic techniques in order to estimate, 
with a physical model, the detection efficiency for a realistic supernova GW signal. 
This signal corresponds to the case of a stiff equation of state, a small initial differential rotation, 
and a moderate rotational kinetic energy \cite{ref:garching}; this leads to 
a waveform with a large negative peak followed by a ring-down phase (regular collapse). 
More up-to-date simulations \cite{ref:dimmelmeier07,ref:dimmelmeier07b} using hydrodynamical 
models with realistic nuclear equations of state tend to confirm the general features of the regular core collapse. 
In this present study we are not interested in the details of the waveform.
The waveform is linearly polarized and the simulations have been done similarly to the Gaussian signals.\\

With the recent breakthroughs in the field of numerical relativity (NR),
many groups have been able to simulate the evolution of binary BHs through
three stages: inspiral, merger and ring-down \cite{ref:pretorius,ref:gsfc,ref:jena}.
These waveforms have been shown to be consistent with the already existing analytical
results using post-Newtonian and BH perturbation theories \cite{ref:buonanno06,ref:baker06,ref:berti,ref:hannam}
for the inspiral and ring-down phases respectively. Furthermore, despite the differences in the
numerical methods employed, gauges chosen, and methods adopted to evolve
the systems, qualitatively the results from all groups show good agreement \cite{ref:agreement}. 
All these provide a motivation to use these results from
a data analysis perspective. Many such attempts have already been made to search the three phases of the binary 
black hole evolution using NR outcomes \cite{ref:ajith,ref:pan,ref:buonanno06}. 
We apply these results from the point of view of a GW burst data analysis
and use the NR waveforms to assess the efficiency of the EGC burst pipeline to
detect the BBH merger.  
For this application we used the numerical relativity (NR) simulations of the non-spinning, equal mass
BBHs provided by the Goddard Space Flight Center group \cite{ref:gsfc}. 
The injected waveforms were GW strains from the leading $l=2,m=2$ spin-weighted spherical 
harmonics from the simulations.  
These waveforms had (approximately) 3 cycles of inspiral, followed by merger, and then a ring-down.

Unlike the ad-hoc waveforms or the linearly polarized case of SN waveforms,
one has to include the effect of the orbital inclination angle $\iota$ 
for the BBH merger waveforms.
Eq.~(\ref{eq:antennapattern-adhoc}) should be rewritten in order to  include the effect of orbital inclination as
\begin{eqnarray}
\label{eq:h_forbinary}
h(t)=F_{+}(\theta,\phi,\psi)\,A_+(\iota)\,h_+(t;\iota=0)+F_{\times}(\theta,\phi,\psi)\,A_{\times}(\iota)\,h_{\times}(t;\iota=0),
\end{eqnarray}
where $F_{+}$ and $F_{\times}$ are the usual antenna pattern functions, $A_{+}=-\frac{1}{2}(1+\cos^2 \iota)$,
$A_{\times}=-\cos \iota$ and
$h_+(t;\iota=0)$ and $h_{\times}(t;\iota=0)$ are the polarizations for the situation of zero inclination angle $\iota$.
In this case, the simulated waveforms are generated by randomizing all four angles involved.
The NR waveforms for the BBH mergers in the total mass range 5$M_\odot$--150$M_\odot$ were injected into the C7 data
in order to calibrate the detection efficiency of the pipeline to these types of mergers.
The range of masses chosen for the injection was decided in accordance with the sensitivity of the C7 run.

\subsection{Detection efficiency of the GW burst search with the software injections}
\label{sec:deteff}
\begin{figure}[t]
\begin{center}
\epsfig{file=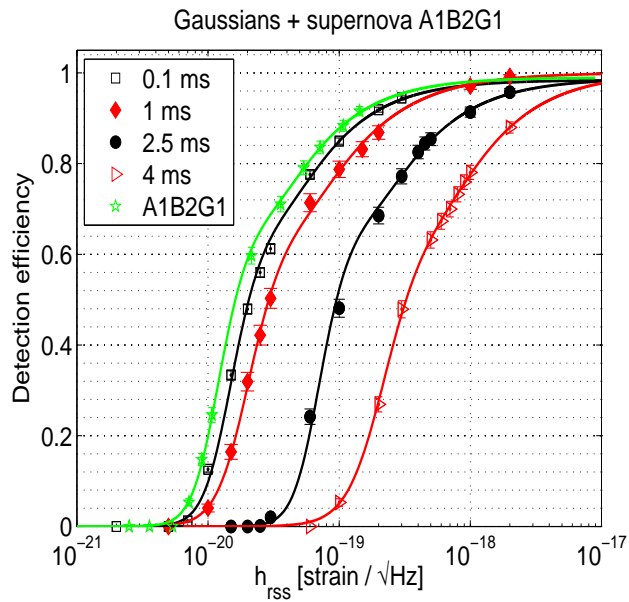, width=9cm, height=8cm}
\caption{\label{fig:eff_GA} Detection efficiency of the search analysis as a function of the signal strength $h_{rss}$ at 
Earth for the Gaussian waveforms (with $\tau$ specified in the legend) and one core collapse simulated waveform (A1B2G1).
The efficiencies were computed for sources at random sky locations. The error bars take into account only the statistical error 
on the detection efficiency.}
\end{center}
\end{figure}

\begin{figure}[t]
\begin{center}
\begin{tabular}{cc}
\epsfig{file=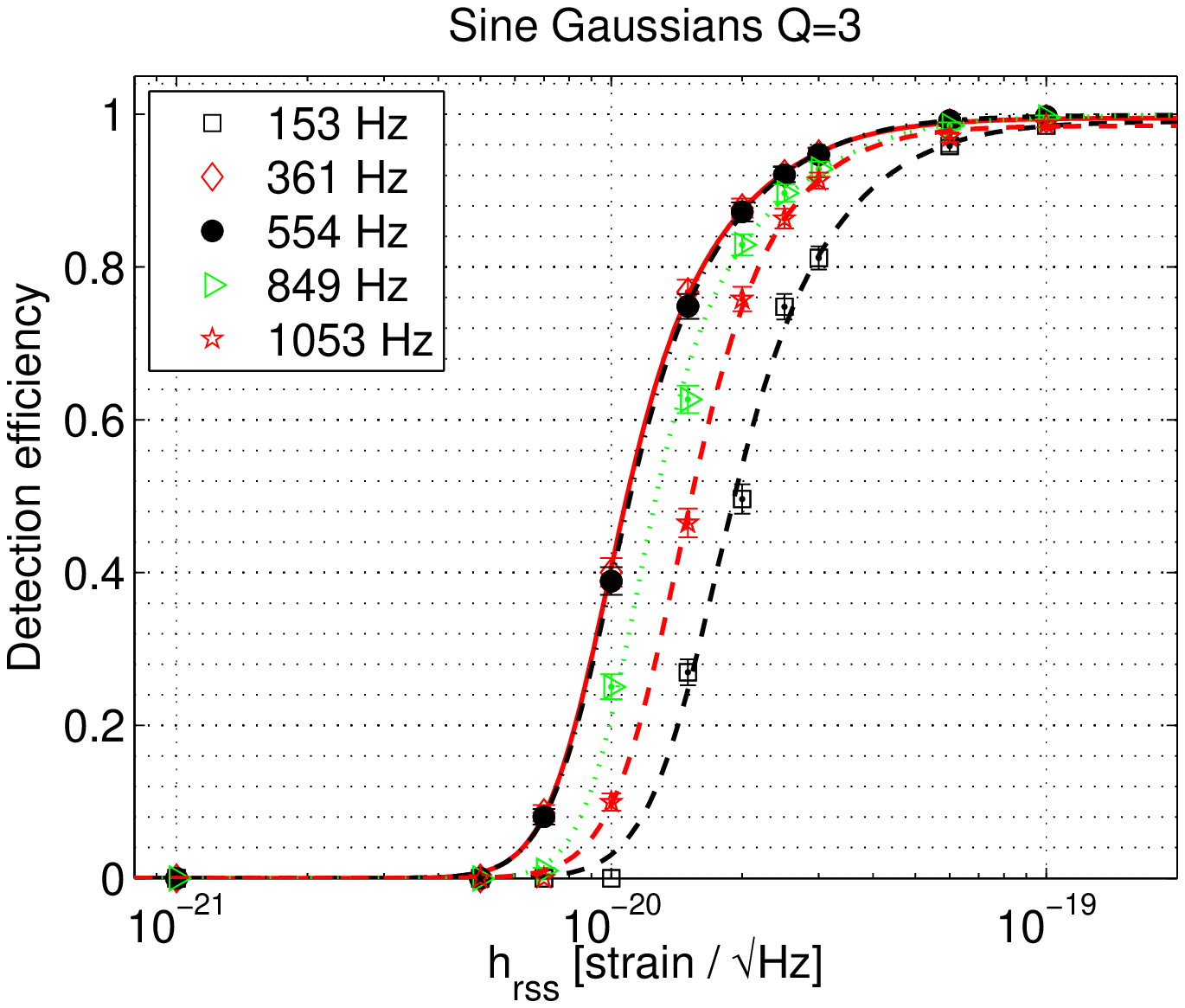, width=9cm, height=8cm} &
\epsfig{file=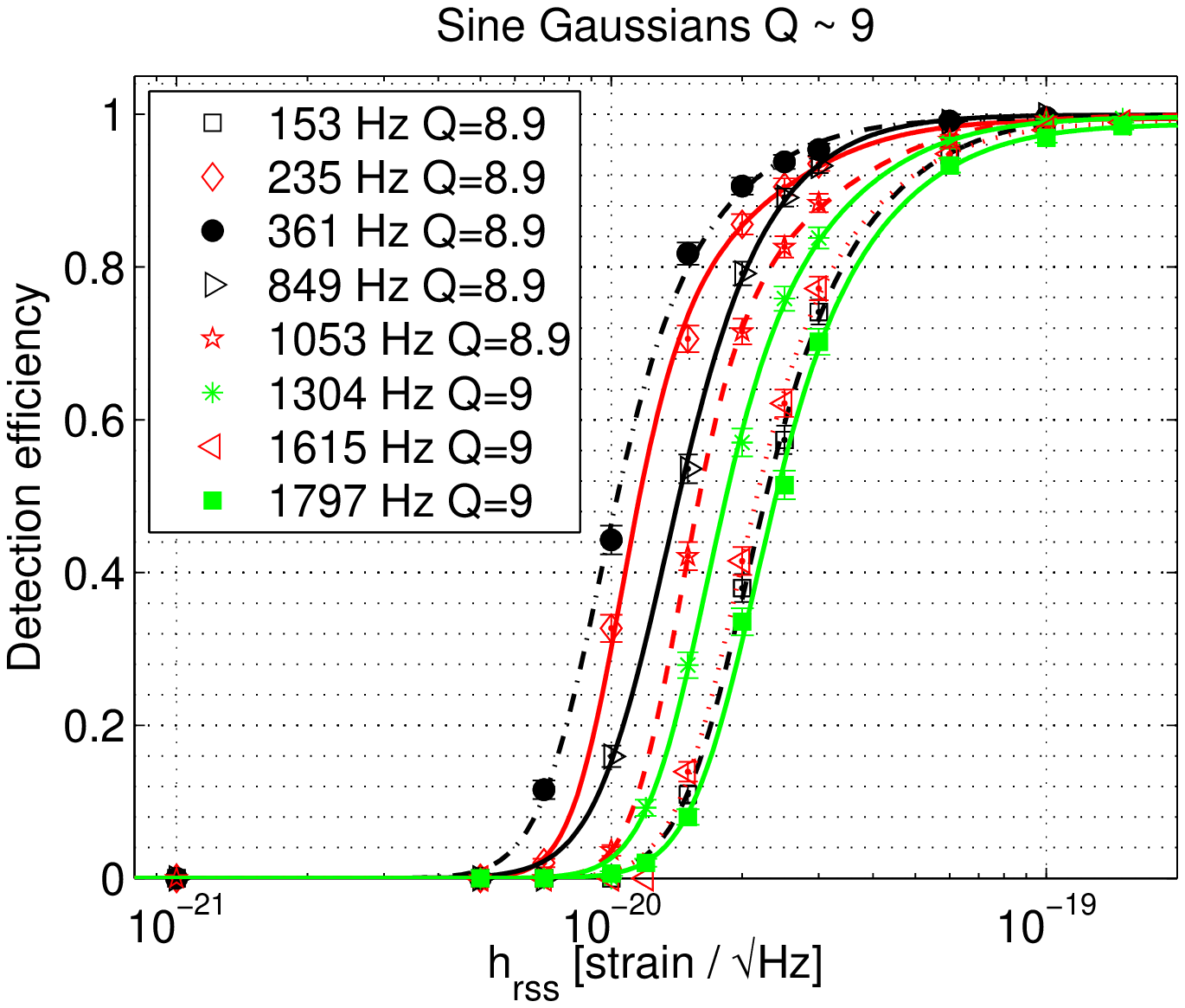, width=9cm, height=8cm}\\
\end{tabular}
\caption{\label{fig:eff_SG} Detection efficiency of the search analysis as a function of the signal strength $h_{rss}$ at 
Earth for Sine Gaussian waveforms with different central frequencies $f_0$ and for two different values of 
the $Q_0$ factor: left $Q_0$ = 3, right $Q_0 \sim $ 9. The efficiencies were computed for sources at random sky locations.
The error bars take into account only the statistical error 
on the detection efficiency.}
\end{center}
\end{figure}

\begin{figure}[t]
\begin{center}
\epsfig{file=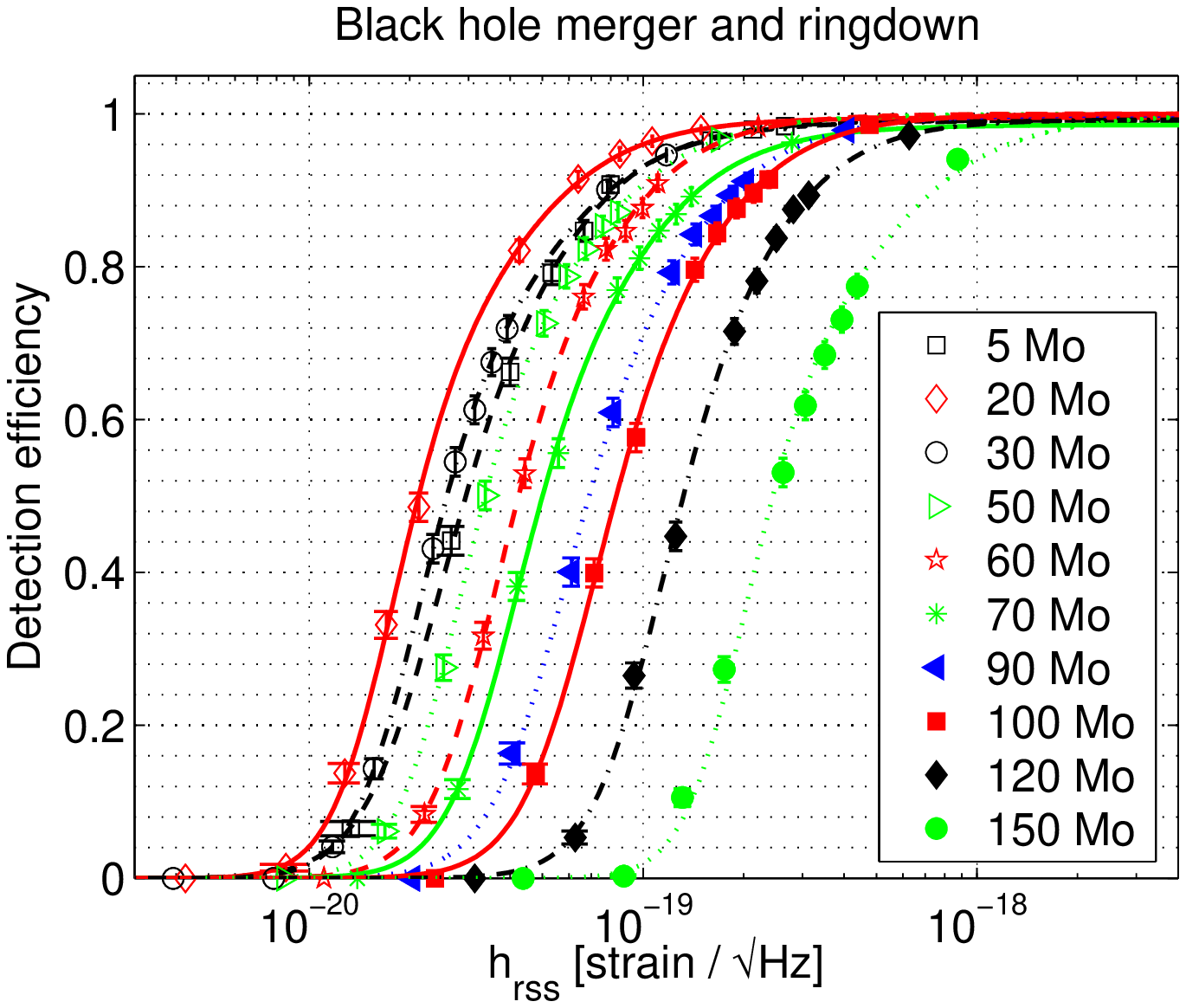, width=9cm, height=8cm}
\caption{\label{fig:eff_NR} Detection efficiencies from the search analysis as a function of the signal strength $h_{rss}$ 
for the BBH merger and ring-down waveforms for different BH total masses (but where the two
individual component masses are equal). The efficiencies were 
computed for random polarization, inclination angles, and sky positions.
The error bars take into account only the statistical error 
on the detection efficiency.}
\end{center}
\end{figure}

We applied the same data quality criteria (DQ and vetoes) on the trigger list containing the waveform injections.
The detection efficiency is defined as the fraction of injected signals reconstructed with a SNR larger than 11.3 corresponding 
to the loudest event found in the C7 data triggers.
As the veto dead time is rather large, the vetoed periods have not been considered when computing the detection efficiency. 
FIGs. \ref{fig:eff_GA}, \ref{fig:eff_SG} and \ref{fig:eff_NR} show the detection efficiency as a function of $h_{rss}$ for 
each family of waveforms.
The error on the efficiency data points takes into account only the statistical error.
To estimate the values of $h_{rss}$ at efficiency 50\% and 90\% the four parameters of an asymmetric sigmoid function have been fitted 
to the data points.
The efficiency $\epsilon (h_{rss})$ is defined as follows, 
\begin{equation}
\epsilon(h_{rss}) = \frac{\epsilon^{max}}{1+{r^{mid}}^{\alpha(1+\beta \tanh(r^{mid}))}},
\end{equation}
where $\epsilon_{max}$ is the maximal efficiency obtained for strong signals (it should tend to unity).
$r^{mid}$ is the ratio $h_{rss} / h_{rss}^{mid}$ where $h_{rss}^{mid}$ is the strain amplitude at half height i.e., 
we have $\epsilon(h_{rss}^{mid})=\epsilon^{max}/2$. 
$\alpha$ and $\beta$ are respectively the slope and the asymmetry of the sigmoid function. 
The parameters were estimated by minimizing a $\chi ^2$ function.
$\epsilon^{max}$ is constrained to remain smaller than unity.
For signals for which $\epsilon^{max}$ does not reach unity, we checked that this is due to a fraction of signals 
with an injected SNR smaller than the threshold of 11.3.
This is especially the case for the Gaussian and the A1B2G1 core collapse simulated waveforms, as expected.
Indeed, as the frequency content of the Gaussian signals is maximal at low frequency, for a fixed $h_{rss}$ and a 
random sky position, we expect a non negligible fraction of low SNR Gaussian signals.

As expected for the Gaussian signals, the performance decreases as the width $\tau$ of the Gaussian increases.
For $\tau$ greater than 4 ms, the detection efficiency is already quite poor, as shown in FIG. \ref{fig:eff_GA}.
Note that the results obtained for the supernova signal A1B2G1 are very similar to the Gaussian with $\tau$=0.25 ms 
(see TABLE \ref{tab:eff-GA-DFM});
this is not surprising as the A1B2G1 signal has a large amplitude peak whose width is approximately 0.35 ms.
FIG. \ref{fig:eff_SG} shows the detection efficiency of the analysis obtained with the Sine Gaussian waveforms;
the performance follows the trend of the C7 Virgo noise floor shown in FIG. \ref{fig:sensitivity}, which is expected
for a signal well localized in frequency, such as a Sine Gaussian. 

The performance obtained for the BBH merger waveforms for a total mass M from 5 $M_{\odot}$ up to 150 $M_{\odot}$ (displayed in FIG. 
\ref{fig:eff_NR}) were not as good as for the Sine Gaussian signals.
The best $h_{rss}^{50\%}$ was achieved for M=20 $M_{\odot}$ with $h_{rss}^{50\%} \simeq 2.2 \times 10^{-20} / \sqrt{Hz}$.
That was at least two times worse than the best Sine Gaussian result.
The merger frequency can be approximated by the empirical formula \cite{ref:baker,ref:LIGO_burst_S4}, deduced from FIG. 8 
in \cite{ref:baker06b},
\begin{equation}
\label{eq:fmerger}
f_{\rm merger} \simeq  \frac{15 \mathrm{kHz}}{M},
\end{equation}
where $M$ is the  mass (in units of solar mass) of the final BH formed after merger.
M is smaller than the sum of the initial BH masses.
For $M$=20 $M_{\odot}$ this corresponds to 750 Hz, which is well inside the frequency range of the search.
On the contrary, for masses above 100 $M_{\odot}$, $f_{\rm merger}$ is smaller than 150 Hz, which is the lower 
frequency boundary of this search.
The EGC pipeline for this kind of waveform is still able to catch the ring-down part of the signal, which has an 
appreciable contribution to the SNR, while the performance
for a Sine Gaussian at the $f_{\rm merger}$ frequency would hardly be detectable.
For the low mass signal the drop of performance is explained by the fact that the frequency at the merger exceeds the upper frequency
boundary of this search, and moreover, the equivalent $Q$ is expected to be larger than 16 (the upper value chosen in this search).

The value of $h_{rss}$ corresponding to 50\% and 90\% of efficiency ($h_{rss}^{50\%}$ and $h_{rss}^{90\%}$)
have been determined using the fitted sigmoid functions and are given in TABLES \ref{tab:eff-SG}, \ref{tab:eff-GA-DFM} 
and \ref{tab:eff-BHBH} for Sine Gaussians, 
Gaussians, Supernova, and BBH merger signals respectively.
The best value of $h_{rss}^{50\%}$, $1.05 \times 10^{-20} / \sqrt{Hz}$, was obtained with the Sine Gaussian of frequency
 $f_0$=361 Hz and $Q_0$ = 9; this signal is 15 times higher than the noise floor at this frequency.
Note that bad performance obtained for the 100 Hz Sine Gaussian is expected since the lower frequency range of the search was 
150 Hz. Moreover the C7 sensitivity is quite poor below 150 Hz.

The quality of the fitting procedure has been controlled by checking that the $\chi^2$ of the fitting function remains close to 
unity within at maximum a factor 3. 
Increasing the error on the efficiency by up to a factor 3 does not change the result on $h_{rss}$ to within 1\%.
The statistical error on $h_{rss}^{50\%}$ and $h_{rss}^{90\%}$ have been computed using the covariance matrix of the efficiency curves;
for the Sine Gaussian and BBH waveforms the error on $h_{rss}^{50\%}$ was below 1.3\%. Note that for $h_{rss}^{90\%}$ the errors 
were higher, between 6\% and 13\%.
For the Gaussian and core collapse waveforms the uncertainty is on average larger, indicating that the 
quality of the fitting was worse. The errors are between 1\% and 4\% for all the Gaussians, except for the Gaussian with $\tau$=4 ms 
for which the error on $h_{rss}^{50\%}$ reaches 11\%.
However, the biggest uncertainty on the $h_{rss}^{50,90\%}$ values comes from the error on the GW strain ampliude calibration, 
which is of 40\%.
Including the systematic error due to the calibration uncertainty leads to an error on $h_{rss}^{50,90\%}$ between 
40\% and 43\%.\\

\begin{table}[t]
\begin{tabular}{ccccccc}
\hline
           & \multicolumn{3}{c}{50\% efficiency} & \multicolumn{3}{c}{90\% efficiency}\\
\hline
$f_0$ (Hz) & $~~~~ Q=3 ~~~~$ & $~~~~ Q=8.9~ \rm{or} ~ 9 ~~~~$ & $~~~~ Q=3 ~~~~$ & $~~~~ Q=8.9 ~{\rm or} ~ 9 ~~~~$  \\
\hline

100    & 4.5   & 10.6     &  9.1   & 21.7    \\
153    & 1.9   & 2.3      &  3.9   & 4.6     \\
235    & 1.2   & 1.2      &  2.3   & 2.5     \\
361    & 1.1   & 1.0      &  2.2   & 2.0     \\
554    & 1.1   & 1.1      &  2.3   & 2.3     \\
849    & 1.3   & 1.4      &  2.6   & 2.6     \\
945    & -     & 1.5(*)   &  -     & 2.9(*)  \\
1053   & 1.5   & 1.6(*)   &  2.8   & 3.6(*)  \\
1172   & -     & 1.8(*)   &  -     & 3.6(*)  \\
1304   & -     & 1.8(*)   &  -     & 3.8(*)  \\
1451   & -     & 1.9(*)   &  -     & 3.7(*)  \\
1615   & -     & 2.2(*)   &  -     & 4.5(*)  \\
1797   & -     & 2.4(*)   &  -     & 5.1(*)  \\

\hline
\end{tabular}
\caption{\label{tab:eff-SG} $h_{rss}$ corresponding to 50\% and 90\% efficiency in units of $10^{-20}$ strain $/\sqrt{Hz}$ 
obtained for the ad-hoc Sine Gaussian waveforms. 
The efficiencies were averaged over the full C7 data set for random sky positions. The symbol - means that no simulated
waveforms were available. The symbol (*) indicates that we used simulated waveform with $Q$=9 instead of 8.9.}
\end{table}

\begin{table}[t]
\begin{tabular}{ccc}
\hline
$\tau$ (ms)  & 50\% efficiency & 90\% efficiency\\
\hline
0.1          & 2.1     &     15.9  \\
0.25         & 1.8     &     11.6  \\
0.5          & 1.8     &     12.7  \\
1.0          & 2.9     &     24.4  \\
2.5          & 9.6     &     79.0  \\
4.0          & 32.4     &     242.  \\
\hline
A1B2G1       &  1.63    &      12.4 \\
\hline
\end{tabular}
\caption{\label{tab:eff-GA-DFM}$h_{rss}$ corresponding to 50\% and 90\% efficiency in units of $10^{-20}$ strain $/\sqrt{Hz}$ 
for the Gaussian waveforms of different widths and one core collapse simulated waveform (A1B2G1). 
The efficiencies were averaged over random sky position. }
\end{table}

\begin{table}[t]
\begin{tabular}{ccc}
\hline
total mass ($M_{\odot}$)  & 50\% efficiency & 90\% efficiency\\
\hline
5            & 3.0     &      8.2   \\
10           & 2.3     &      5.6   \\
20           & 2.1     &      6.0   \\
30           & 2.6     &      7.9   \\
40           & 3.0     &      8.0   \\
50           & 3.4     &      9.6   \\
60           & 4.2     &      10.7   \\
70           & 5.0     &      14.6   \\
80           & 5.8     &      14.8   \\
90           & 6.9     &      18.7   \\
100          & 8.3     &      21.3   \\
120          & 13.2    &      32.4   \\
150          & 24.5    &      74.4   \\
\hline
\end{tabular}
\caption{\label{tab:eff-BHBH}$h_{rss}$ corresponding to 50\% and 90\% efficiency in units of $10^{-20}$ strain $/\sqrt{Hz}$ for 
the BBH numerical relativity waveforms. 
The total mass of the system varies from 5 $M_{\odot}$ up to 150 $M_{\odot}$. }
\end{table}

\section {Search results}
\label{sec:section6}
\subsection{Event rate and GW strength upper limits}

\begin{figure}[t]
\begin{center}
\begin{tabular}{cc}
\epsfig{file=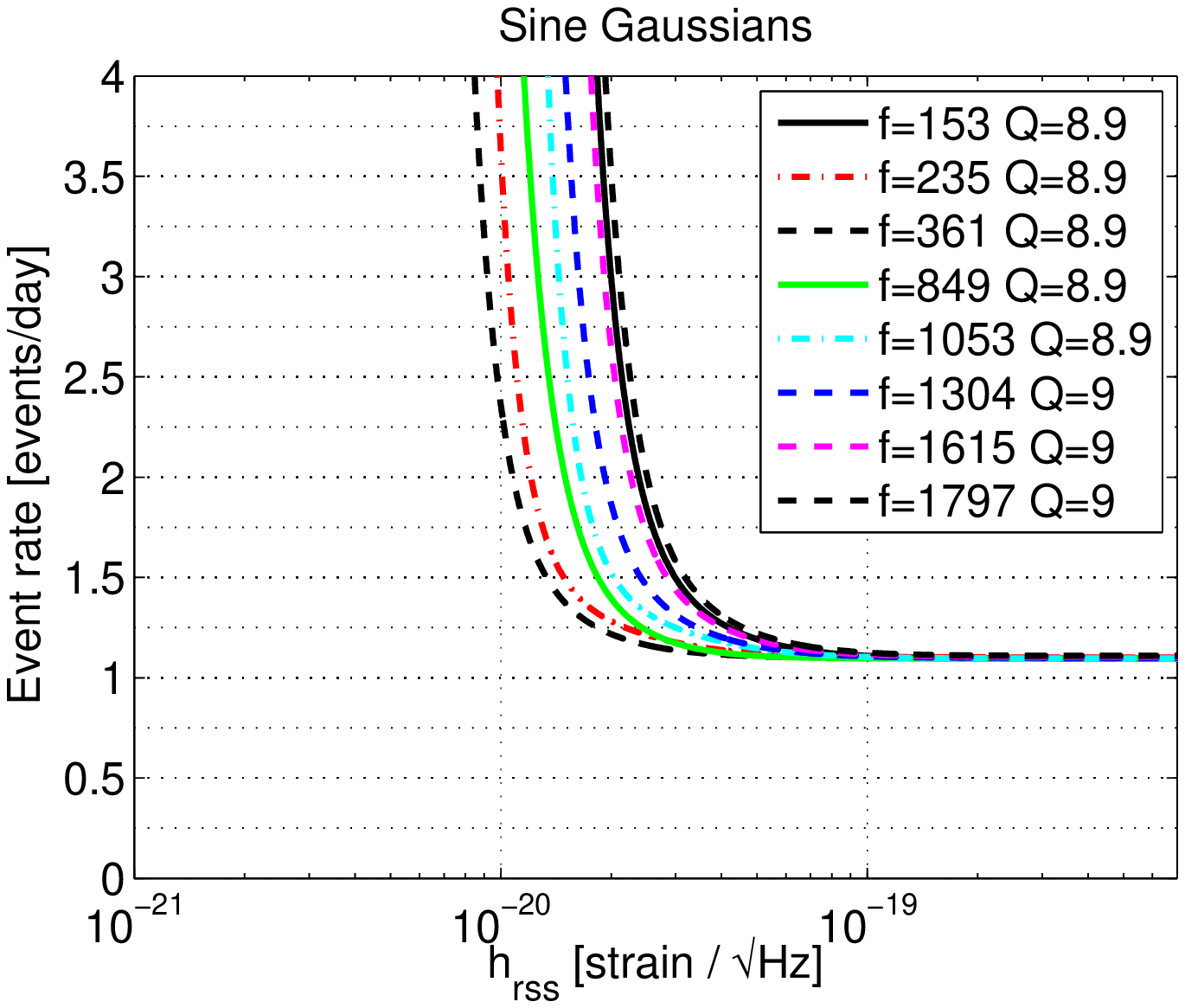, width=9cm, height=8cm} &
\epsfig{file=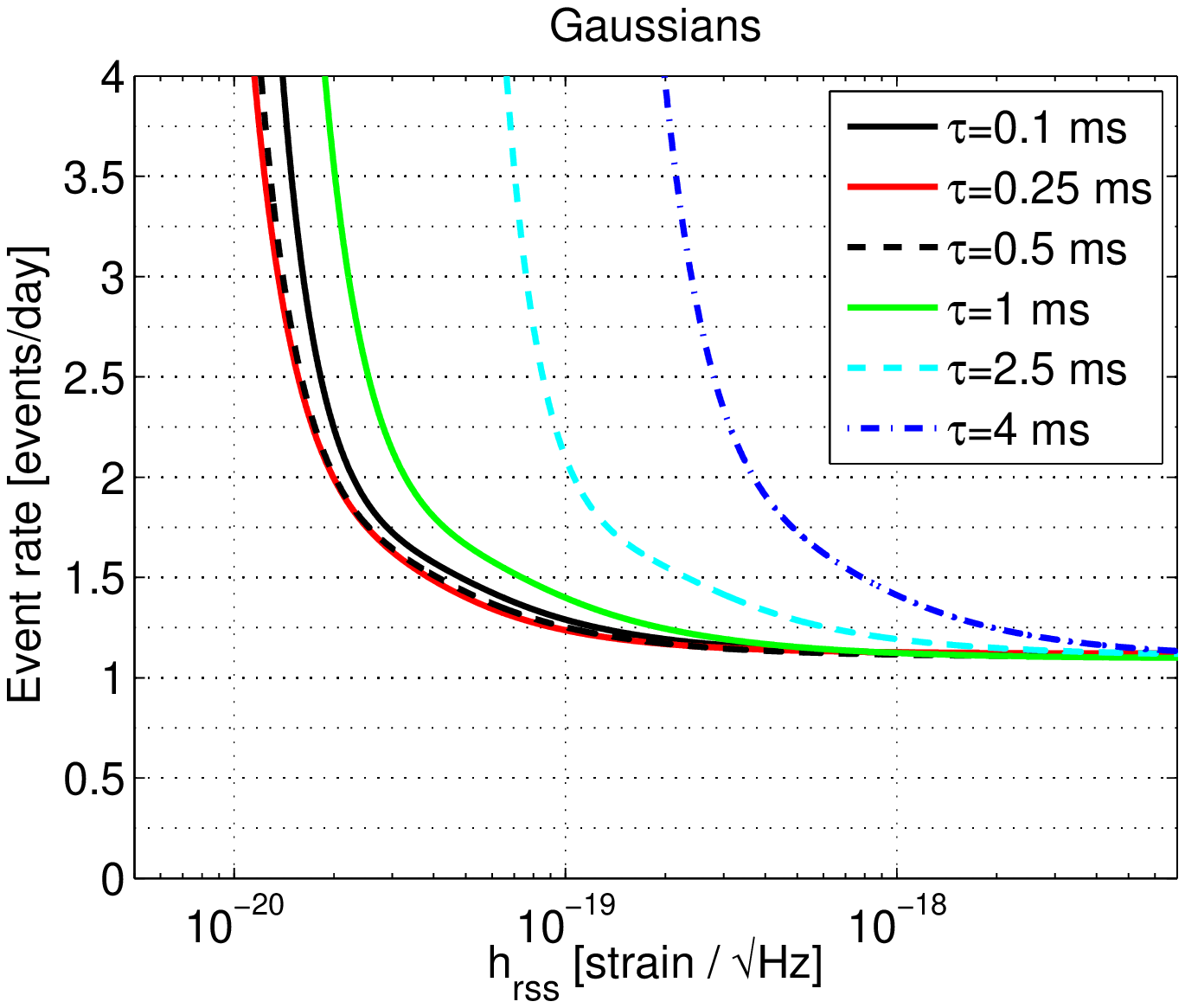, width=9cm, height=8cm}\\
\end{tabular}
\caption{\label{fig:UL} Event rate as a function of the $h_{rss}$ exclusion plots at 90\% confidence level obtained in the Virgo C7 data.
The left plot corresponds to GW burst events that can be modeled by a Sine Gaussian with a Q of 8.9 or 9 for frequencies, $f$, covering the 
most promising part of the C7 frequency range. The right plot corresponds to GW burst events that can be modeled by a peak of different widths, 
$\tau$, from 0.1 ms up to 4 ms.}
\end{center}
\end{figure}

From Section \ref{sec:background} we can assert that we have not observed any GW event with a SNR above 11.3, corresponding 
to the loudest event SNR.
Knowing the detection efficiency as a function of the signal strength for different kinds of signals, we can now derive a 
90\% confidence level (CL) upper limit on the event rate that depends on the strength of the signal.
Using the loudest event statistic as described in \cite{ref:LES}, the 90\% CL exclusion on the event rate of a signal detected with an efficiency of $\epsilon(h_{rss})$ 
is given by,
\begin{equation}
{\cal R}^{90\%} (h_{rss}) = \frac{-ln(1-0.9)}{T \, \epsilon(h_{rss})},
\label{eq:UL}
\end{equation}
where $T \simeq 2.1$ days is the effective observation time for the C7 run (veto times have been subtracted from the total analyzed time).
The results are given in FIG. \ref{fig:UL} for the Sine Gaussian waveforms (Q $\sim$ 9) and the Gaussian waveforms.

As expected for strong signals, the exclusion rate reaches the asymptotic value of $\frac{-ln(1-0.9)}{T} \simeq 1.1$ events per day 
whatever the signal type.
We can exclude at 90\% confidence level a GW event rate ${\cal{R}}_{90\%}$, assuming 100\% detection efficiency, of 1.1 events per day 
for $T \simeq 2.1$ days of observation.
On the other hand, depending on the characteristics of the signals the exclusion rate, when the signal becomes weak, is rather 
different.
The vertical asymptotic value in FIG. \ref{fig:UL} gives an indication of how the sensitivity of the search depends on a given frequency 
for the Sine Gaussian, or on the width of the Gaussian waveform.
The exclusion results for Sine Gaussian waveforms with Q=3 (from 153 Hz up to 1053 Hz only) are very similar to the Sine Gaussian 
results with Q=9.

\subsection{Astrophysical interpretation}
\label{sec:astro}
In the previous section we determined the sensitivity of the C7 data analysis to different potential GW burst 
sources in terms of $h_{rss}$.
This can be translated into an astrophysical estimate on the amount of GW energy emitted by a source or 
the distance reach, depending on the type of sources we consider.
For simplicity we have assumed in our ad-hoc waveform simulations that the emission is isotropic,
although this may not be the case for a real astrophysical signal.
In the case of a Sine Gaussian waveform with a central frequency $f_0$ 
and quality factor $Q\gg1$, one obtains for the GW energy emitted by the source \cite{ref:ShapiroTeukolsky,ref:riles}
\begin{equation}
\label{eq:EGWSG}
E_{\rm GW} \simeq \frac{r^2\,c^3}{4\,G} \, (2\pi\,f_0)^2\,h_{\rm rss}^2,
\end{equation}
where $r$ is the (non-cosmological) distance to the source.
Similarly, for a Gaussian waveform, the emitted GW energy is
\begin{equation}\label{eq:EGWG}
E_{\rm GW} \simeq \frac{r^2\,c^3}{4\,G}\,\frac{1}{\tau^2}\,h_{\rm rss}^2,
\end{equation}
where $\tau$ is the width of the Gaussian.

The above expressions can be used to derive an astrophysical sensitivity with the present C7 data 
analysis in different ways.
For sources at a fixed distance, an estimate on the emitted GW energy can be obtained by using the smallest value of
$h_{rss}^{50\%}$.
Conversely, if we know the energy emitted in GWs, this information can be used to infer the maximum distance up to 
which such an event could have been observed by Virgo with an efficiency larger than 50\%.
In what follows we discuss these various possibilities, quantifying the detectability of supernova events and 
BBH merger events where the nature of waveforms and the energy of emission are known through the numerical simulations.

\subsubsection{Detectability of an arbitrary burst GW event close to the galactic center}

Considering the Sine Gaussian and Gaussian waveforms that give the best results in terms of $h_{rss}^{50\%}$, we can compare
the estimation of the maximal GW energy emitted by a hypothetical GW burst source in the vicinity of the galactic center,
assuming a distance of 10 kpc.
With the 361 Hz Sine Gaussian we obtained $h_{rss}^{50\%}$ of $1.05\times10^{-20}/{\sqrt {\rm Hz}}$, which
translates into an energy of $\sim 3\times10^{-5}M_\odot~ c^2$. 
For a Gaussian waveform, the most favorable case corresponds to $\tau=0.25$ ms and $h_{rss}^{50\%} \simeq 1.79\times10^{-20}/{\sqrt {\rm Hz}}$; the associated energy is $\sim 2.8\times10^{-4}M_\odot~ c^2$. 
Hence the most optimistic limit on the energy differs almost by an order of magnitude between 
the Gaussian and Sine Gaussian waveforms, while the detection performance in terms of $h_{rss}$ is only two times worse for a 
Gaussian with respect to a Sine Gaussian.

\subsubsection{Detectability of supernova events}

We now consider the predictions on the GW energy produced by the collapse of a stellar iron core to form 
a neutron star. 
It is assumed that the progenitor star rotates fast enough to generate sufficient deformation.
We do not consider here the GWs generated after the core collapse by the development of an oscillation 
of the proto-neutron star lasting a few hundred of milliseconds
\cite{ref:ott06} since we did not expand the search for corresponding high Q oscillating signals. 
A few hundred of milliseconds at a frequency around 700 Hz corresponds to a Q of a few hundred.
Using the core collapse waveform A1B2G1 injected into the data to determine the efficiency for that particular signal, 
we can derive the distance reach corresponding to 50\% of efficiency as we do for the BBH waveforms; 
the distance is 0.15 kpc.
However, this waveform might not be representative of the overall core collapse bounce waveforms. 
Indeed, depending on the progenitor model parameters (mass, rotation rate, equation of state) the GW energy 
prediction can vary by several orders of magnitude \cite{ref:dimmelmeier07b}: 
$3 \times 10^{-10} M_{\odot}~ c^2 < E_{GW} < 7 \times 10^{-8} M_{\odot}~ c^2$ with peak frequency which can be as high 
as 800 Hz.
However, these recent simulations \cite{ref:dimmelmeier07b} seem to indicate that the waveforms are rather generic regardless 
of the dynamic of the collapse (pressure dominated or centrifugal bounce). 
Even in the case of centrifugal bounce, the waveforms exhibit a positive pre-bounce followed by a negative peak, but 
the typical frequency is lower than in the case of a uniform differential rotation for which the typical
frequency is around 700 Hz.  
For instance, if one considers the model with a 20 $M_{\odot}$ progenitor mass, a moderate differential rotation value (A=1000 km) 
and a moderate initial rotation rate ($\beta$ = 0.5\%), $E_{GW} \simeq 2.32 \times 10^{-8} M_{\odot}$ and the characteristic 
frequency is 715 Hz using the equation of state provided by Shen \cite{ref:shen}.
We should not specifically assume that supernova GW signals are monochromatic waveforms like Sine Gaussians. 
They do not look like Gaussians either (they are more localized in frequency than Gaussian).
Taking the $h_{rss}^{50\%}$ value for the $\tau$ = 0.5 ms Gaussian (that roughly corresponds to the width 
of the peak of the considered waveform), we obtain a maximal distance range of 
0.18 kpc.
Using the Sine Gaussian expression, and using the $h_{rss}^{50\%}$ for a central frequency 715 Hz we obtain 
a distance of 0.12 kpc.
For the A1B2G1 waveform used in this paper, and considering a radiated energy of
$E_{GW} \simeq 2.13 \times 10^{-8} M_{\odot} c^2$ computed using Eq. (5) of \cite{ref:dimmelmeier02}, 
we obtain a distance of 0.12 kpc whatever the assumption 
made on the waveform (Gaussian of $\tau$ = .35 ms or a Sine Gaussian of $f_0$=650 Hz).
These numbers indicate the order of magnitude of the maximal detection distance we have reached during 
Virgo's C7 run.
This achieved distance is somewhat small, and we expect to gain at least one order of magnitude 
with Virgo data acquired during more recent observation runs.

\subsubsection{Distance reach for binary black hole mergers}
\begin{figure}[h]
\begin{center}
\epsfig{file=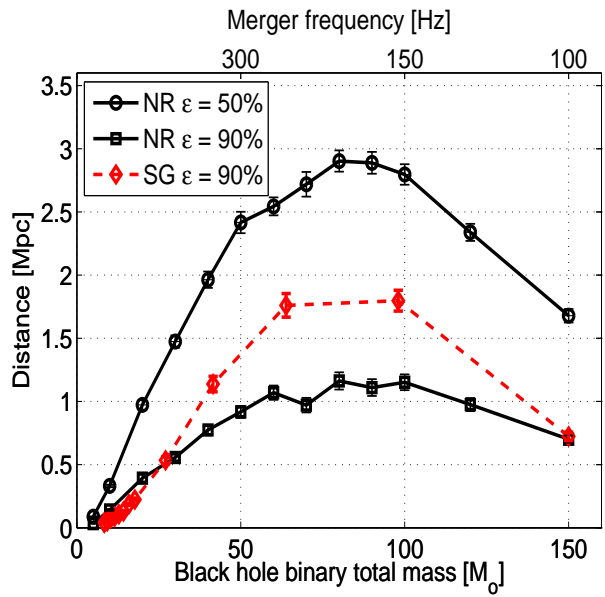, width=9cm, height=8cm}
\caption{\label{fig:BHBH-distance} 
Detection distance of the black hole binary source with efficiency of 90\% and 50\% as a function of the total mass of the 
system. 
The curves with circles and squares have been obtained by estimating the EGC pipeline detection efficiency to numerical 
relativity black hole binary waveforms. 
The curve with diamonds was derived from the results obtained with Sine Gaussian waveforms 
assuming that the waveform when the two black holes merge resembles a Sine Gaussian. 
The error bars take into account only the statistical error on the detection efficiency.}
\end{center} 
\end{figure}

From the detection efficiency results obtained with the BBH waveforms in Section \ref{sec:deteff}, 
one can derive a distance at which EGC is able to detect a signal 
with 50\% or 90\% efficiency, respectively $d^{50\%}$ and $d^{90\%}$.
These values have been extracted from the results of the fits of the detection efficiency as a function of the distances 
for which the BBH sources have been simulated. 
An asymmetric sigmoid function has been used and the errors on $d^{50\%}$ and $d^{90\%}$ have been estimated as they have 
for $h_{rss}^{50\%}$ and $h_{rss}^{90\%}$. 
FIG. \ref{fig:BHBH-distance} shows the $d^{50\%}$ and $d^{90\%}$ detection distances as a function of the total mass of 
the system. 
$d^{90\%}$ gives an indication of the average distance range up to which this search could have been able to 
detect a BBH source.
With the present choice of threshold on the SNR, our search is sensitive to BBH mergers in the mass range
20 $M_\odot$-150 $M_\odot$ up to a distance of 1-3 Mpc (0.5-1Mpc) for 50\% (90\%) detection efficiency. 
The distance reach is maximum for a BBH of total mass 80-90 $M_\odot$. 
In the C7 sensitivity band, this mass range corresponds to signals where the merger part is dominant.
Besides, for a mass higher than 100 $M_\odot$, the merger frequency (given by Eq. (\ref{eq:fmerger})) is below the frequency 
lower bound studied in the analysis (150 Hz).
This fact adds to the degraded C7 run sensitivity below 150 Hz, and explains why the distance reach for the high mass 
waveforms
is decreasing.

This bound on the distance can, in principle, be translated into an upper limit on the event rate, 
assuming a Gaussian distribution of BBH masses around 80 $M_{\odot}$.
This analysis would require combining a model for the galaxies, their blue light luminosity distribution as a function
of distance and mass, along with the antenna pattern of the detector; this has been proposed in 
\cite{ref:Galaxy07} for the LIGO detectors.
But given the limited volume of the universe that the present search is capable of sampling, 
we postpone this more complete analysis until we analyze the most recent Virgo data with its improved sensitivity.
Moreover, the results have been obtained only for non-spinning equal mass binary systems.

An alternative way of deriving the distance reach for BBH mergers is to assume that the merger and ring-down phases 
of the BBH waveforms resemble Sine Gaussians and use the results obtained from the Sine Gaussian waveforms to derive the 
detection distance.
In order to use Sine Gaussians to model the BBH mergers, we approximate the NR waveforms to be Sine Gaussians with
peak frequency $f_0$ equal to the frequency at the merger ($f_{\rm merger}$) of the NR waveforms using 
Eq. (\ref{eq:fmerger}) following \cite{ref:LIGO_burst_S4}. 
It should be noted that this definition of $f_{\rm merger}$ is only approximate. 
Besides, it is not obvious how to assign a precise value of $Q$ to the Sine Gaussian that approximates a BBH waveform.
However, taking into account the typical duration $\tau$ of the ``merger'' part of the waveform and using the 
relation, exact for a Sine Gaussian, $\tau = Q / (2 \pi f_0)$, one could deduce that the Q is between 10 and 20 for 
the high mass BBH and as large as 40 for the low masses. 
This is larger than the boundary of Sine Gaussian templates bank used in this analysis. 
Hence, we may suspect that for the low mass ($< 20 M_\odot$) the EGC pipeline
is not optimally reconstructing the SNR of the NR waveforms.
However, it is straightforward to show that $h_{rss}^{50\%}$, the quantity which is used in the derivation, does 
not depend on $Q$ for Sine Gaussians as long as $Q$ is sufficiently large (TABLE \ref{tab:eff-SG} confirms this).
For all these reasons, in the discussion that follows, we consider a Sine Gaussian with Q $\sim$ 9.
In order to determine the distance reach one should make an assumption on the total energy emitted by the source 
and on the source emission pattern. 
According to NR simulations \cite{ref:baker06b}, $\sim$ 3.5\% of the total rest mass energy would be radiated away 
as GWs.
Following \cite{ref:LIGO_burst_S4}, we can use Eq.~(\ref{eq:EGWSG}) and the values of $h_{rss}$ in 
TABLE \ref{tab:eff-SG} to derive the distances up to which such BBH merger events could be observed with
an efficiency of 50\% or 90\%.
The results for the distance reach at an efficiency of 90\% are presented in
FIG.\,\ref{fig:BHBH-distance} together with the distance reach obtained using NR waveforms at 
the same detection efficiency. At intermediate masses and frequencies the distance
reach obtained when BBH mergers are modeled with SG waveforms is about 60\%
larger than the one obtained when more realistic NR waveforms are used. 
A detailed comparison of the two waveforms shows that this difference is 
partly due to the broader spectral distribution of the NR waveform for which 
part of the signal energy falls out of the detector and analysis bandwidth. The remaining 
difference is due to the EGC pipeline, which by construction matches more 
efficiently with the SG waveforms.

\section{Conclusion}
\label{conclusion}
In this paper, we have reported on a search for short duration GW bursts in the Virgo C7 data from within the 
150Hz - 2 kHz frequency band;
this corresponds to the best sensitivity achieved by Virgo during the C7 run.
This search, carried out using the output of only one GW detector, required first the understanding and then the possibility
of vetoing all identified sources of noise.
Special care has been taken to assure ourselves that the procedure does not suppress a GW event, and that 
the dead time remains small.
Unfortunately, the non stationary data features prevent a search during vetoed time periods, which amounts to an accumulated 
dead time of 20.2 \%.
Though this rather large dead time limits the astrophysical interest of the search, it has been possible to calculate
a sensitivity for various GW burst sources.
The best sensitivity that has been reached in terms of the square root of the strain amplitude at 50\% of efficiency, 
$h_{rss}^{50\%}$, was 1.1 $\times 10^{-20} ~ / \sqrt{\rm{Hz}}$ at 361 Hz, a frequency region where Virgo's sensitivity was optimal.
If one considers the whole set of waveforms studied in this paper, the sensitivity lies between 
$10^{-20}$ and $10^{-19} / \sqrt{\rm{Hz}}$.
However, one should note that the error on the sensitivity can be as large as 40\% due to uncertainties in the calibration
of the strain amplitude.
Those numbers can be compared to GW burst searches carried out with LIGO detectors during the S2 data taking in 
2003 \cite{ref:LIGO_burst_S2}.
Indeed, the C7 Virgo sensitivity is comparable to that of 4-km LIGO detectors during their S2 run;
Virgo during C7 was a little bit more sensitive above 500 Hz and a bit less below.
Not surprisingly, the search sensitivity obtained with C7 Virgo data is similar to LIGO S2 results for the same kind of 
waveforms: $h_{rss}^{50\%}$ having values  between $10^{-20}$ and $10^{-19} / \sqrt{\rm{Hz}}$.

Besides the search of un-modeled GW burst waveforms, we have shown using recently computed numerical relativity waveforms 
that burst pipelines can be used efficiently to search for BBH merger GW emission.
In the case of non-spinning equal mass system, a distance reach of 2.9 Mpc for 80 $M_{\odot}$ total mass has been inferred from 
the C7 data assuming an efficiency of 50\%.
This mass range corresponds to when the BH merger part of the signal dominates over the inspiral part.
That also corresponds to a frequency region where ground base interferometers achieve their best sensitivity.
It is interesting to note that the sensitivity in the merger dominated frequency range is up by roughly a factor 
of 10 in the most recent Virgo data (acquired in the later half of 2007). 
This would correspond to a distance reach up to 30 Mpc (12 Mpc) at 50\% (90\%) efficiency.
This distance reach improvement is not enough to tackle regions where the number of galaxies becomes interesting
given the expected BBH event rate. Further upgrades, like Advanced Virgo, may be needed to have a reasonable
probability of detection.
Despite that, and given the improvement of the Virgo sensitivity in the low frequency region, we will expand the EGC 
pipeline parameter space to explore more efficiently from this region. 
Though matched filtering with a bank of templates, including inspiral, merger and ring-down phases, might have a better 
detection efficiency for BBH mergers, it is important to complement the search with other techniques 
not based on tailored matched filtering but using generic templates, such as Sine Gaussians.
This approach could provide a method that would be more robust at finding increasingly complicated signals 
that depend on a larger number of astrophysical parameters from the sources, such as a black hole's spin in the case 
of a black hole - neutron star system.

Finally, one should add that the recent Virgo data taking (VSR1) will be analyzed in coincidence with LIGO S5 data.
That is especially interesting in the high frequency region ($>$ 600 Hz) where the sensitivity of the four detectors 
are all limited by shot noise.
This network of four non-aligned interferometers with comparable sensitivity should be able to detect GW signal emitted
by a supernova anywhere in the galaxy. 

\begin{acknowledgments}
The authors thank H. Dimmelmeier for fruitful discussions and providing the new core collapse waveforms.
We thank the numerical relativity group at NASA Goddard Space Flight Center for sharing the results of their binary black 
hole simulations with us.
We also want to thank B. Kelly for valuable discussions about these waveforms.
\end{acknowledgments}


\begin{thebibliography}{}
\label{sec:biblio}
\bibitem{ref:virgo_recent} F. Acernese {\it et al} (Virgo Collaboration), Class. Quantum Grav. {\bf 24}, S381 (2007).
\bibitem{ref:C7_CB} F. Acernese {\it et al} (Virgo Collaboration), Class. Quantum Grav. {\bf 24}, 5767 (2007).
\bibitem{ref:C7_pulsar} F. Acernese {\it et al} (Virgo Collaboration), Class. Quantum Grav. {\bf 24}, S491 (2007).
\bibitem{ref:C7_bars} F. Acernese {\it et al} (Virgo, ROG and AURIGA Collaborations), Class. Quantum Grav. {\bf 25}, 205007 (2008).
\bibitem{ref:C7_GRB} F. Acernese {\it et al} (Virgo Collaboration) Class. Quantum Grav. {\bf 25}, 225001 (2008).
\bibitem{ref:zwerger} T. Zwerger and E. Mueller, Astron. Astrophys, {\bf 320}, 209 (1997).
\bibitem{ref:dimmelmeier02} H. Dimmelmeier, J. A. Font and E. Mueller, Astron. Astrophys, {\bf 393}, 523 (2002).
\bibitem{ref:ott04} C. D. Ott {\it et al}, Astrophys. J., {\bf 600}, 834 (2004). 
\bibitem{ref:shibata} M. Shibata and Y. I. Sekiguchi, Phys. Rev. D {\bf 69}, 084024 (2004).
\bibitem{ref:ott06} C. D. Ott {\it et al}, Phys. Rev. Lett., {\bf 96}, 201102 (2006). 
\bibitem{ref:flanagan} E. E. Flanagan and S. A. Hughes, Phys. Rev. D {\bf 57}, 4535 (1998).
\bibitem{ref:baker} J. G. Baker {\it et al}, Phys. Rev. D {\bf 73}, 104002 (2006).
\bibitem{ref:pretorius} F. Pretorius, Phys. Rev. Lett. {\bf 95}, 121101 (2005).
\bibitem{ref:campanelli} M. Campanelli {\it et al}, Phys. Rev. Lett. {\bf 96}, 111101 (2006).
\bibitem{ref:kokkotas} K. D. Kokkotas and B. G. Schmidt, "Quasi-Normal Modes of Stars and Black Holes", Living Rev. Relativity {\bf 2} (1999), \url{http://www.livingreviews.org/lrr-1999-2}.
\bibitem{ref:meszaros} P. Meszaros, Rept. Prog. Phys. {\bf 69}, 2259-2322 (2006).
\bibitem{ref:ferrari} V. Ferrari, G. Miniutti and J. A. Pons, Class. Quantum Grav. {\bf 20}, S841 (2003).
\bibitem{ref:damour} T. Damour, A. Vilenkin, Phys. Rev. D {\bf 71}, 063510 (2005).
\bibitem{ref:postnov06} K. A. Postnov and L. R. Yngelson, "The Evolution of Compact Binary Star Systems", Living Rev. Relativity {\bf 6} (2006), \url{http://www.livingreviews.org/lrr-2006-6}.
\bibitem{ref:oshaughnessy05} R. O'Shaughnessy, C. Kim, T. Frakgos, V. Kalogera, K. Belczynski, Astrophys. J. {\bf 633}, 1076, (2005).
\bibitem{ref:LIGO_burst_S2} B. Abbott {\it et al} (LSC), Phys. Rev. D {\bf 72}, 062001 (2005).
\bibitem{ref:pretorius07} F. Pretorius, "Relativistic Objects in Compact Binaries: From Birth to Coalescence", Editor: Colpi et al. Pulisher: Springer Verlag, Canopus Publishing Limited, arXiv:0710.1338 (2007).
\bibitem{ref:LIGO_burst_S1} B. Abbott {\it et al} (LSC), Phys. Rev. D {\bf 69}, 102001 (2004).
\bibitem{ref:LIGO_burst_S4} B. Abbott {\it et al} (LSC), Class. Quantum Grav. {\bf 24}, 5343-5369 (2007).
\bibitem{ref:IGEC_burst} P. Astone {\it et al}, Phys. Rev. D {\bf 68}, 022001 (2003).
\bibitem{ref:TAMA_burst} M. Ando {\it et al} (TAMA Collaboration), Phys. Rev. D {\bf 71}, 082002 (2005).
\bibitem{ref:goddard} J. G. Baker {\it et al}, Phys. Rev. D {\bf 75}, 123024 (2007).
\bibitem{ref:virgo} F. Acernese {\it et al} (Virgo Collaboration), ``The Virgo detector'', to be submitted to NIM.
\bibitem{ref:SA} S. Braccini {\it et al}, (Virgo Collaboration), Astropart. Phys. {\bf 23}, 557 (2005)
\bibitem{ref:locking} F. Acernese {\it et al} (Virgo Collaboration), Astropart. Phys. {\bf 30}, 29-38 (2008)
\bibitem{ref:hrec_C7} L. Rolland, {\it private communication}
\bibitem{ref:EP} W. G. Anderson {\it et al}, Phys. Rev. D {\bf 63}, 042003 (2001).
\bibitem{ref:WB} S. Klimenko, G. Mitselmakher, Class. Quantum Grav. {\bf 21}, S1819 (2004).
\bibitem{ref:PF} G. M. Guidi {\it et al}, Class. Quantum Grav. {\bf 21}, S815 (2004). 
\bibitem{ref:Q} S. Chatterji {\it et al}, Class. Quantum Grav. {\bf 21}, S1809 (2004).
\bibitem{ref:EGC} A.-C. Clapson {\it et al}, Class. Quantum Grav. {\bf 25}, 035002 (2008).
\bibitem{ref:arnaud03} N. Arnaud {\it et al}, Phys. Rev. D {\bf 67}, 102003 (2003).
\bibitem{ref:HK} J. Hoshen, R. Kokelman, Phys. Rev. B {\bf 14}, 3438 (1976).
\bibitem{ref:LV_Burst} F. Beauville {\it et al}, Class. Quantum Grav. {\bf 25}, 045002 (2008).
\bibitem{ref:MF} N. Arnaud {\it et al}, Phys. Rev. D {\bf 67}, 062004 (2003).
\bibitem{ref:suspension} F. Acernese {\it et al} (Virgo Collaboration), Class. Quantum Grav. {\bf 21}, S425 (2004).
\bibitem{ref:gouaty} R. Flaminio, R. Gouaty, E. Tournefier, Virgo technical document, VIR-NOT-LAP-1390-313.
\bibitem{ref:vetoes} N. Christensen, P. Shawhan, G. Gonz\'{a}lez (for the LIGO Scientific Collaboration), Class. Quantum Grav. {\bf 21}, S1747 (2004).
\bibitem{ref:PQMon} K. Koetter, I. Heng, M. Hewitson, K. Strain, G. Woan and H. Ward, Class. Quantum Grav. {\bf 20} S895-S902 (2003).
\bibitem{ref:Hanna} C. R. Hanna (for the LSC Collaboration), Class. Quantum Grav. {\bf 23} S17-S22 (2006).
\bibitem{ref:Christensen05} N. Christensen (for the LSC Collaboration), Class. Quantum Grav. {\bf 22} S1059-S1068 (2005).
\bibitem{ref:siesta} B. Caron {\it et al}, Astropart. Phys. {\bf 10}, 369 (1999).
\bibitem{ref:gsfc} J. G. Baker {\it et al}, Phys. Rev. D {\bf  73}, 104002 (2006).
\bibitem{ref:garching} \url{http://www.mpa-garching.mpg.de/rel\_hydro/}
\bibitem{ref:BP} N. Christensen, Phys. Rev. D, {\bf 46}, 5250 (1992).
\bibitem{ref:dimmelmeier07} H. Dimmelmeier {\it et al}, Phys. Rev. Lett. {\bf 98}, 251101 (2007).
\bibitem{ref:dimmelmeier07b} H. Dimmelmeier {\it et al}, Phys. Rev. D {\bf 78}, 064056 (2008).
\bibitem{ref:jena} B. Bruegmann {\it et al}, Phys. Rev. Lett. {\bf 92}, 211101 (2004).
\bibitem{ref:buonanno06} A. Buonanno, G. B. Cook, F. Pretorius, Phys. Rev. D {\bf 75}, 124018 (2007). 
\bibitem{ref:baker06} J. G. Baker {\it et al}, Phys. Rev. Lett. {\bf 99}, 181101 (2007).
\bibitem{ref:berti} E. Berti {\it et al}, Phys. Rev. D {\bf 76}, 064034 (2007).
\bibitem{ref:hannam} M. Hannam {\it et al}, Phys. Rev. D {\bf 77}, 044020, (2008).
\bibitem{ref:agreement} J. G. Baker {\it et al}, Class. Quantum Grav. {\bf 24}, S25-S31 (2007).
\bibitem{ref:ajith} P. Ajith {\it et al}, Class. Quantum Grav. {\bf 24} S689-S700, (2007).
\bibitem{ref:pan} Y. Pan {\it et al}, Phys. Rev. D {\bf 77}, 024014, (2008).
\bibitem{ref:baker06b} J. G. Baker {\it et al}, Phys. Rev. {\bf D} 73, 104002 (2006).
\bibitem{ref:LES} P. R. Brady, J. D. E. Creighton, and A. G. Wiseman, Class. Quant. Grav. {\bf 21}, S1775 (2004).
\bibitem{ref:ShapiroTeukolsky} S. L. Shapiro and S. A. Teukolsky, ``BlackHoles, White Dwarfs and Neutron Stars'', John Wiley a\& Sons, New-York (1983).
\bibitem{ref:riles} K. Riles, LIGO technical document, LIGO-T040055-00-Z.
\bibitem{ref:shen} H. Shen {\it et al}, Prog. Theor. Phys {\bf 100}, 1013 (1998).
\bibitem{ref:Galaxy07} R. K. Kopparapu {\it et al}, Astrophys. J. {\bf 675}, 1459, (2008).
\end{thebibliography}
\end{document}